\documentclass[5p]{elsarticle}         
\biboptions{sort&compress}

\usepackage{lineno,hyperref}
\usepackage{epstopdf} 
\usepackage{url} 
\usepackage{subfigure}

\modulolinenumbers[5]

\journal{Nuclear Instrument and Methods in Physics Research A}

\begin{document}

\begin{frontmatter}

\title{Performance test of the MAIKo active target}

\author[kyoto]{T.~Furuno \corref{cor1} \fnref{fn1}}
\ead{furuno@rcnp.osaka-u.ac.jp}

\author[kyoto]{T.~Kawabata \fnref{fn2}}
\author[rcnp]{H.~J.~Ong}
\author[rcnp]{S.~Adachi}
\author[lbnl]{Y.~Ayyad}
\author[kyoto]{T.~Baba}
\author[kyoto]{Y.~Fujikawa}
\author[ibs]{T.~Hashimoto}
\author[kyoto]{K.~Inaba}
\author[kyoto]{Y.~Ishii}
\author[tokai]{S.~Kabuki}
\author[kyoto]{H.~Kubo}
\author[cyric]{Y.~Matsuda}
\author[kyoto]{Y.~Matsuoka}
\author[kyoto]{T.~Mizumoto}
\author[kyoto]{T.~Morimoto}
\author[kyoto]{M.~Murata}
\author[kyoto]{T.~Sawano}
\author[rcnp]{T.~Suzuki}
\author[kyoto]{A.~Takada}
\author[tud]{J.~Tanaka}
\author[rcnp,beihang]{I.~Tanihata}
\author[kyoto]{T.~Tanimori}
\author[rcnp,vietnam]{D.~T.~Tran}
\author[kyoto]{M.~Tsumura}
\author[kyoto]{H.~D.~Watanabe}

\address[kyoto]{Department of Physics, Kyoto University, Kitashirakawa Oiwake, Sakyo, Kyoto 606-8502, Japan}

\address[rcnp]{Research Center for Nuclear Physics, Osaka University,
               10-1 Mihogaoka, Ibaraki, Osaka 567-0047, Japan}

\address[lbnl]{Lawrence Berkeley National Laboratory, 1 Cyclotron Road, Berkeley, CA 94720, USA}

\address[ibs]{Rare Isotope Science Project, Institute for Basic Science, 
               Yuseong-daero 1689-gil, Yuseong-gu, Daejeon 305-811, Korea}

\address[tokai]{School of Medicine, Tokai University, 143 Shimokasuya, Isehara, Kanagawa 259-1193, Japan}

\address[cyric]{Cyclotron and Radioisotope Center,
                Tohoku University, 6-3 Aoba Aramaki, Aoba, Sendai, Miyagi 980-8578, Japan}

\address[tud]{Institut f{\"u}r Kernphysik, Technische Universit{\"a}t Darmstadt,
              Karolinenplatz 5, 64289 Darmstadt, Germany}

\address[beihang]{International Research Center for Nuclei and Particles in the Cosmos,
              and School of Physics and Nuclear Energy Engineering, Beihang University,
              37 Xueyuan Road, Haidian, Beijing 100191, China}

\address[vietnam]{Institute of Physics, Vietnam Academy of Science and Technology, 
                   10 DaoTan, BaDinh, Hanoi 100000, Vietnam}





\cortext[cor1]{Corresponding author}

\fntext[fn1]{Present address: Research Center for Nuclear Physics, Osaka University, 
             10-1 Mihogaoka, Ibaraki, Osaka 567-0047, Japan. 
             Tel.: +81-6-6879-8857;
             Fax: +81-6-6879-8899.}

\fntext[fn2]{Present address: Department of Physics, Osaka University, 
             1-1 Machikaneyama, Toyonaka, Osaka 560-0043, Japan}






\begin{abstract}
  A new active target named MAIKo 
  (Mu-PIC based Active target for Inverse Kinematics$_{\circ}$) 
  has been developed at Kyoto University and Research Center for Nuclear Physics (RCNP),
  Osaka University.
  MAIKo is suited for missing-mass spectroscopy of unstable nuclei at forward scattering angles
  in inverse kinematics.
  MAIKo consists of a time projection chamber (TPC),
  which incorporates a micro-pixel chamber ($\mu$-PIC) as the electron multiplication
  and collection system. 
  In MAIKo, the medium gas also plays the role of
  a reaction target, 
  thus allowing detection of low-energy recoil particles with high
  position resolution.
  The MAIKo TPC was commissioned with He(93\%)+iso-C$_{4}$H$_{10}$(7\%) and
  He(93\%)+CO$_{2}$(7\%) mixture gasses at 430 hPa.
  The gas gain and the angular resolution of MAIKo were evaluated with an alpha source and
  a $^{4}$He beam at 56 MeV.
  The TPC was stably operated up to 1000-kcps beam intensity.
  A tracking algorithm using the Hough transform method has been developed 
  to analyze scattering events.
  An angular resolution of 1.3$^{\circ}$ was achieved for scattered $^{4}$He particles.
\end{abstract}


\end{frontmatter}


\section{Introduction}
Direct reactions with light ions are useful probes
to investigate the structures of both stable and unstable nuclei owing to the 
relative simplicity of reaction mechanisms \cite{direct}.
The direct reactions are dominant processes at beam energy of several 
tens MeV/u.
Various reactions such as elastic and inelastic scattering,
charge exchange, transfer and knock-out reactions
have been widely employed for spectroscopic studies of stable nuclei mostly with
$p$, $d$, $^{3}$He, and $^{4}$He beams.
In these reactions, measurements at forward angles in the center-of-mass (CM) frame are
especially important because the reaction mechanism is simple and the experimental results
are less ambiguously interpreted.
With the recent developments of new facilities which are capable to provide radio-isotope (RI) beams
\cite{ribf,frib,fair,spiral2}, 
measurements of direct reactions have been extended
away from the stability line.
These experiments have been
performed under the inverse kinematic conditions
where hydrogen or helium target is bombarded with RI beams.
There are two methods to determine the excitation energy of the unstable nuclei:
the invariant-mass spectroscopy and the missing-mass spectroscopy.

The invariant-mass spectroscopy has been widely applied in 
many earlier experiments using relatively high-energy ($>50$ MeV/nucleon)
RI beams, to obtain the
exci\-tation-energy spectra of unstable nuclei by detecting all
fragments emitted from the beam particles.
This method enables 
usage of a thick liquid or solid target which ensures the highest yield
at around zero degree.
However, the application of this technique is limited to measurements in which multiplicity of
the fragments is low.

On the other hand, the missing-mass spectroscopy, 
which requires detection of only the recoil particle from the target,
can be applied regardless of the multiplicity.
However, measurements at forward angles in the CM frame
require the detection of low-energy recoil particles.
For example, the energy of the recoil alpha particle is as low as 
0.5 MeV at $\theta_{\mathrm{CM}}=3^{\circ}$
in the alpha inelastic scattering off $^{10}$C at 75 MeV/u,
which is one of our first priority experiments.
In order to detect such low-energy recoil particles with external detectors, the target
must be extremely thin, at the expense of the luminosity.

One solution to the above-mentioned problem is to store RI beams in a storage ring and use an internal gas-jet
target~\cite{exl}.
This technique allows the use of an extremely thin ($\sim$10 pg/cm$^2$) target while maintaining
the necessary luminosity because the unreacted
beams are injected repeatedly onto the target.
Recently, measurements of proton elastic scattering and alpha inelastic scattering were
successfully performed \cite{exl-ela,exl56ni,exl58ni}.
However, this technique is only applicable to nuclei whose life times are longer than
a few seconds because it takes some time to store and cool RI beams in a storage ring.

The use of a time projection chamber (TPC) as an active target is another solution.
The essential feature of the active target is the use of the detection medium gas of the TPC as 
a target gas.
Typically, helium, hydrogen, deuterium or hydrocarbon gas is used as the detection medium gas.
The TPC enables three-dimensional reconstruction of charged-particle trajectories.
Since the reaction occurs inside the sensitive volume of the TPC,
the detection threshold for the recoil particles can be lowered
and the solid angle for the recoil particles is increased up to almost 4$\pi$.
Moreover, the luminosity can be increased by extending the length of a TPC 
along the beam axis while maintaining a low detection threshold.
The reconstructed trajectory of a recoil particle determines 
the recoil angle and recoil energy, which are used to
calculate the missing mass.
Several active target systems have been developed~\cite{ikar,mstpcnim,mayanim,pattpcnim,cat,attpcnim,ROGER2018126},
and measurements using these detectors have been reported
\cite{mayares1,PhysRevLett.113.032504,PhysRevC.92.024316,BAGCHI2015371,RODRIGUEZTAJES2017246,
      PhysRevC.87.054301,PhysRevC.93.014321,BRADT2018155}.
A comprehensive review on the recent developments of active targets 
is given in Ref.~\cite{BECEIRONOVO2015124}.

Recently, a new active target named MAIKo
(Mu-PIC based Active target for Inverse Kinematics$_{\circ}$) was jointly developed by
Kyoto University and Research Center for Nuclear Physics (RCNP), Osaka University to
perform missing-mass spectroscopies using RI beams at several tens MeV/nucleon at the
exotic nuclei (EN) beam line at RCNP \cite{en1,en2}.
We adopt a strip-type readout instead of a pad-type readout to
achieves the finest readout pitch among the existing active targets
while keeping the number of readout channel small.
The gas gain and the angular resolution of MAIKo were evaluated 
by using an alpha source and a $^{4}$He beam at 56 MeV.

The paper is organized as follows.
The design of the MAIKo active target is described in Section~\ref{design}.
Measurements using an alpha source are reported in Section~\ref{alpha}, followed by results
of the performance test using a $^{4}$He beam in Section~\ref{beam}.
The analysis of scattering events is discussed in Section~\ref{scattering}.
The summary and the future outlook are given in Section~\ref{secsummary}.

\section{Design of MAIKo}
\label{design}

\subsection{Overview}
\begin{figure}[htbp]
  \begin{center}
    \includegraphics[width=85mm]{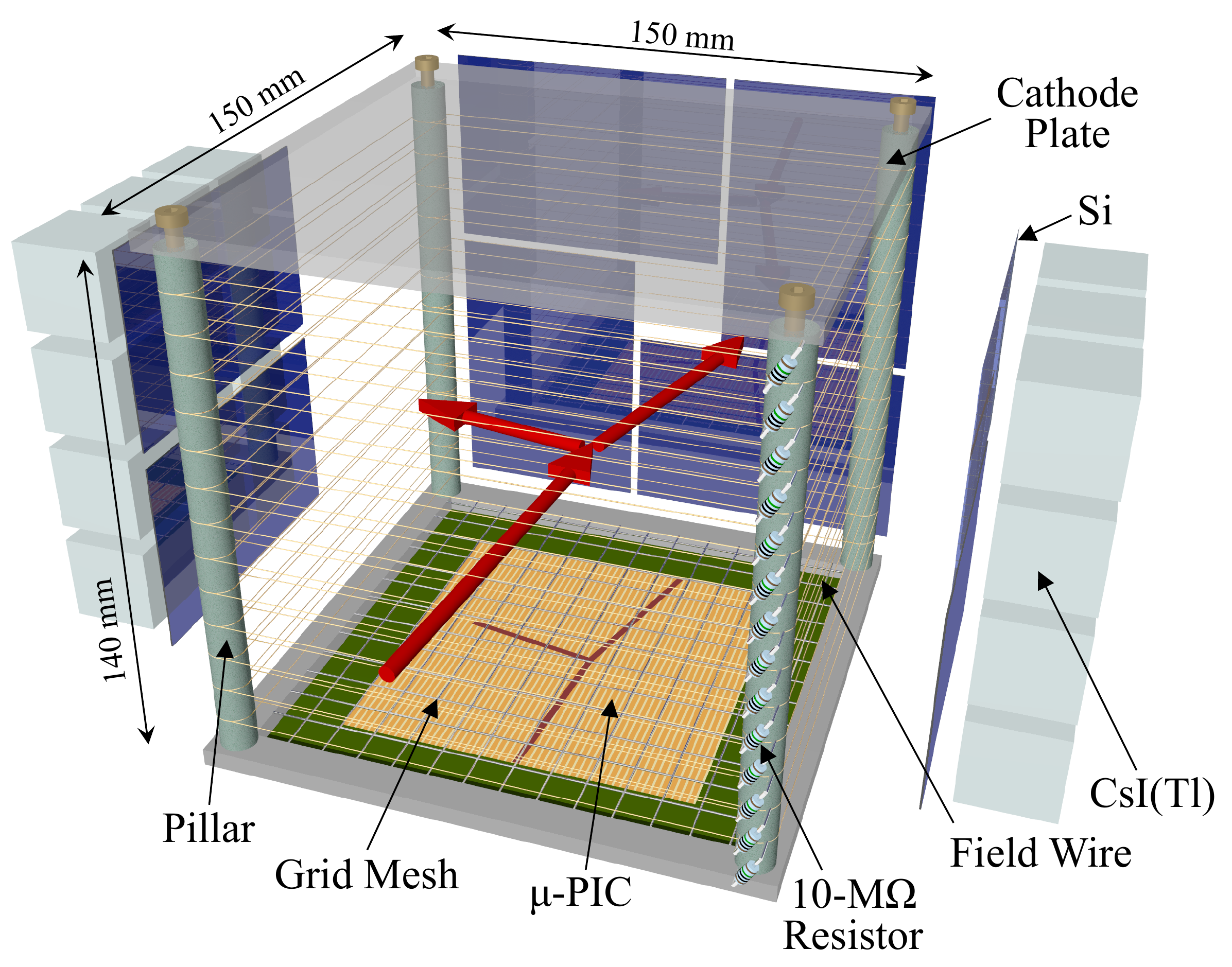}
    \caption{Schematic view of the MAIKo system.}
    \label{at_fig}
  \end{center}
\end{figure}

A schematic view of the MAIKo system is drawn in Fig. \ref{at_fig}.
The field cage of the MAIKo TPC has a volume of 150 $\times$ 150 $\times$ 140 mm$^3$.
The TPC is installed in a stainless-steel vacuum chamber with a
volume of about 30~L.
The chamber is filled with helium gas,
which works as the $^4$He target as well as the detector gas,
with a small fraction of quenching gas.
The pressure of the detector gas can be changed from 100 hPa to 2000 hPa according to the
experimental purpose.

The detector gas is ionized by charged particles,
and the ionization electrons are drifted vertically downwards
along the electric field formed by the TPC field cage.
These electrons are multiplied and detected with the micro-pixel chamber ($\mu$-PIC)~\cite{ochi-mupic}.
The $\mu$-PIC is a micro-pattern gaseous detectors
developed at Kyoto University.
The $\mu$-PIC has been successfully employed in a Compton camera for gamma ray imaging \cite{smile1,smile2,smile3,smile4},
in the dark matter search experiment \cite{MIUCHI201011,doi:10.1093/ptep/ptv041},
and in neutron imaging \cite{PARKER201323,PARKER2013155}.
The sensitive area of the $\mu$-PIC is 102.4 $\times$ 102.4 mm$^{2}$, which limits the sensitive volume of the
TPC to 102.4 $\times$ 102.4 $\times$ 140 mm$^{3}$.
Details of the $\mu$-PIC are described in Section \ref{secupic}.

High-energy recoil particles which punch through the sensitive volume of the TPC 
are detected by ancillary detectors consisting of silicon (Si) detectors 
and thallium-doped cesium iodide [CsI(Tl)] scintillators, 
which are described in Section~\ref{ancillary}.

\subsection{Field cage}
\label{sec_cage}
\begin{figure}[htbp]
  \begin{center}
    \includegraphics[width=85mm]{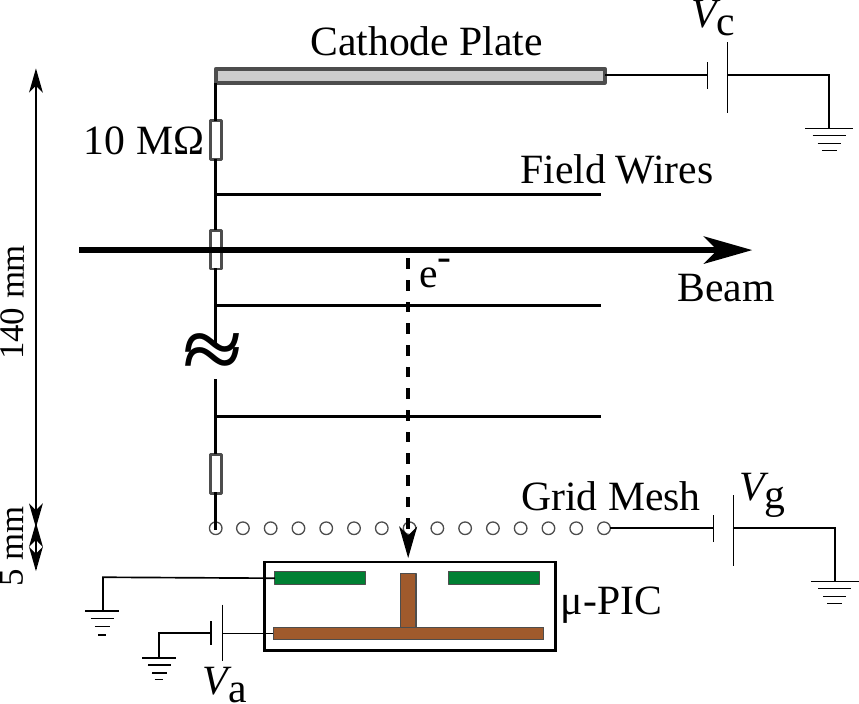}
    \caption{Electric circuit of the field cage.}
    \label{cage}
  \end{center}
\end{figure}

The electric circuit of the field cage is shown in Fig.~\ref{cage}.
The field cage forms a homogeneous vertical electric field inside the sensitive volume of the TPC.
The field-cage cathode is made of a 5-mm-thick aluminum plate.
The size of the cathode plate is 150 $\times$ 150 mm$^2$.
The ionization electrons are drifted towards a nickel grid mesh,
which is placed 140 mm below the cathode plate.
The diameter of the mesh wire is 150 $\mu$m and the pitch size is 0.85 mm (30 wires per inch).
The grid mesh is glued on a glass-reinforced epoxy (G10) frame.
The electric potential of the grid is tuned so that the grid is transparent to the drift electrons but
opaque for the positive ions generated by the avalanche on the $\mu$-PIC.
When the gas pressure is 430 hPa,
negative high voltages of $-3600$ V and $-800$ V are typically applied to the cathode plate
and the grid mesh, respectively, 
thus generating an electric field of 200 V/cm in the field cage.
The ion back flow ratio is estimated to be less than 2$\%$
using a simulation with the Garfield code \cite{garfield}.
Four G10 pillars with a diameter of 10 mm
are mounted between the cathode plate and the grid mesh to
sustain the field-cage structure as well as to hold the field wires.
The drift field is uniformized by 13 Be-Cu wires winding doubly around the G10 pillars
at 10-mm intervals.
The diameter of these field wires is 125 $\mu$m and they are connected via 
a 10-M$\Omega$ metal-film resistor chain to make a voltage divider.
The tension of the wires is kept at 3~N.

A finite element calculation with the computer code neBEM \cite{nebem} was performed 
to check the uniformity of the electric field.
The neBEM code is implemented in the Garfield code.
According to the simulation, the field cage ensures that the distortion of the
electric field is kept within 2$\%$ in the sensitive volume
even when the Si detectors are grounded and installed at 30 mm away 
from the field cage.


\subsection{Micro-pixel chamber ($\mu$-PIC)}
\label{secupic}
\begin{figure}[htbp]
  \begin{center}
    \includegraphics[width=80mm]{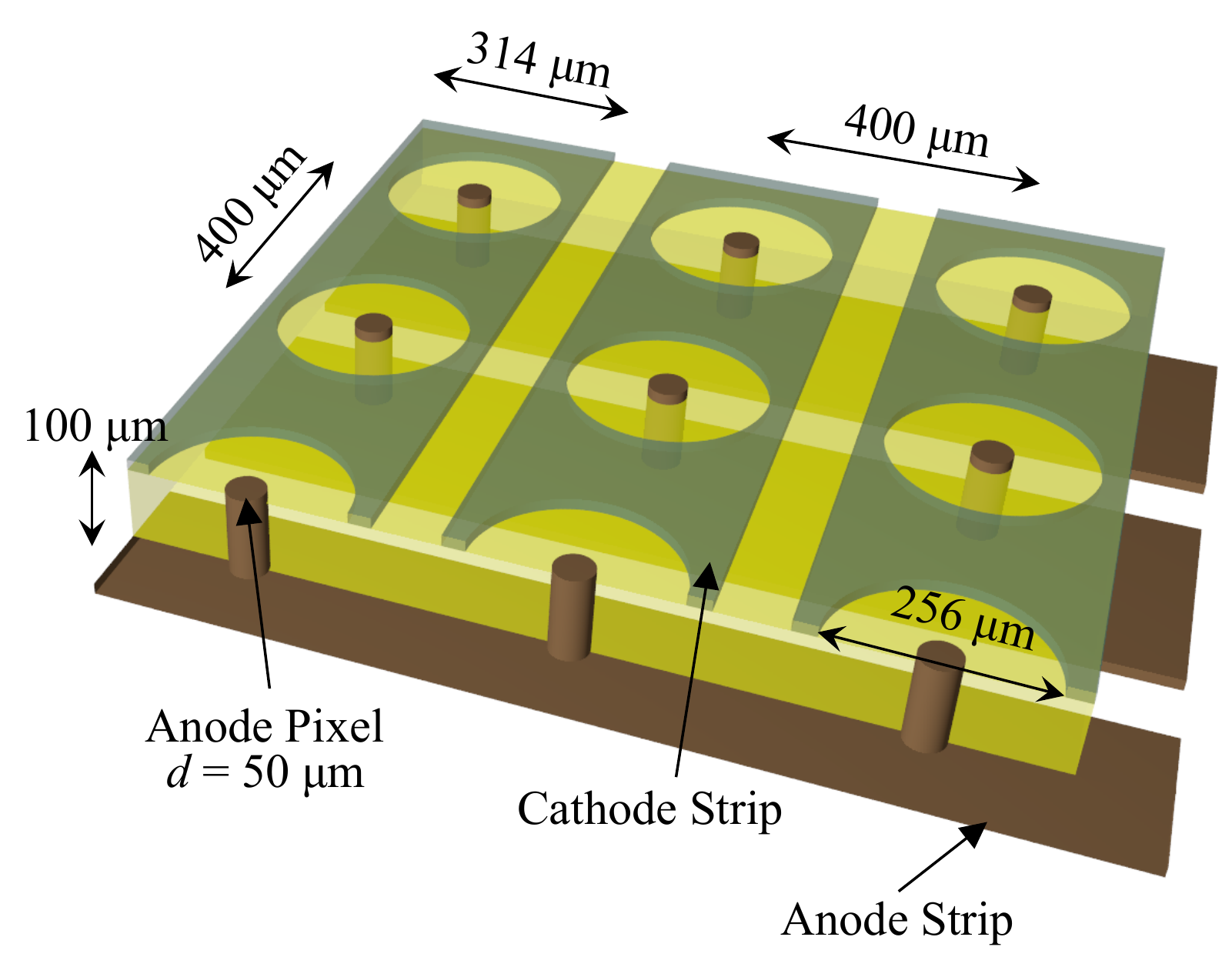}
    \caption{Schematic structure of the $\mu$-PIC \cite{ochi-mupic}.
             The $\mu$-PIC consists of anode pixels and cathode strips fabricated 
             with a pitch of 400 $\mu$m.
             Positive high voltage is applied to the anode pixels while the cathode
             strips are grounded. The drifted electrons form avalanches around
             the anode pixels.
             Signals induced by the electron avalanches are read out by the anode and cathode
             strips which are orthogonally arranged.}
    \label{upic}
  \end{center}
\end{figure}

The drifted electrons are multiplied by the $\mu$-PIC placed 5 mm below the grid mesh.
A schematic structure of the $\mu$-PIC is shown in Fig. \ref{upic}.
The $\mu$-PIC is fabricated using a printing circuit board (PCB) technology.
The anode and cathode electrodes are printed on both sides of a 100-$\mu$m thick
polyimide base.
The cathode strips of 314-$\mu$m width are printed 
on the front side of the base with a pitch of 400 $\mu$m.
These strips have holes of 256-$\mu$m diameter with 400-$\mu$m spacing along the strips.
At the center of the holes, anode pixels are printed.
The diameter of the anode pixels is 50 $\mu$m.
The anode pixels are extended to the back side of the base and connected to the anode
strips which are orthogonal to the cathode strips.
The cathode strips are grounded while positive high voltage is applied to
the anode strips which forms an electric field around the anode pixels 
strong enough to induce an electron avalanche.
Typical values of the high voltage are from 300 to 400 V.

Each strip of the anode and cathode is connected to a readout circuit to
provide a two-dimensional fine image of the ionization electrons.
The $\mu$-PIC consists of the 256 anode strips and the 256 cathode strips 
which determine the sensitive area of 102.4 $\times$ 102.4 mm$^2$.

\subsection{Readout system}
\label{readout}
\begin{figure}[htbp]
  \begin{center}
    \includegraphics[width=80mm]{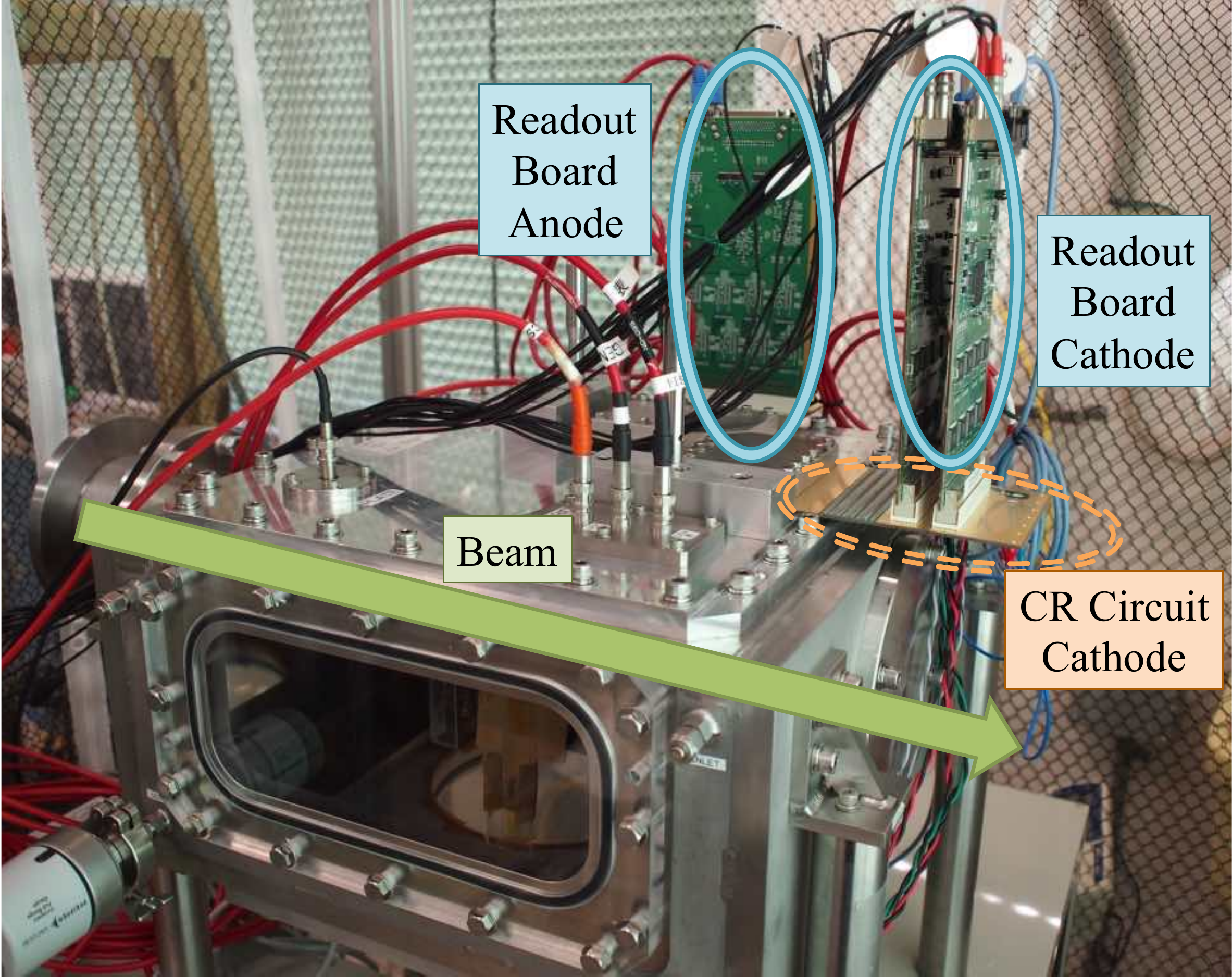}
    \caption{Photograph of the TPC vacuum chamber and the readout boards.
      The CR circuit board for the cathode strips is circled by the orange-dashed lines and
      the readout boards are circled by the blue-solid lines.
      The direction of the beam axis is shown by the green arrow.}
    \label{board}
  \end{center}
\end{figure}

The anode and cathode strips are connected to the capacitor and resistor (CR)
circuit boards.
The CR circuit boards consist of 1-G$\Omega$ resistors for the high voltage bias supply 
and 100-pF capacitors for the coupling to the preamplifiers.
Figure \ref{board} shows a photograph of the CR circuit boards mounted on top of the 
TPC vacuum chamber.
The TPC field cage is installed upside-down in the vacuum chamber.
The PCB board of the $\mu$-PIC is mounted to the lid of the chamber, and the CR 
circuit boards are directly connected to the PCB board.
This CR circuit boards are used also for the vacuum feed-through for the 
$\mu$-PIC signals.

\begin{figure}[htbp]
  \begin{center}
    \includegraphics[width=87mm]{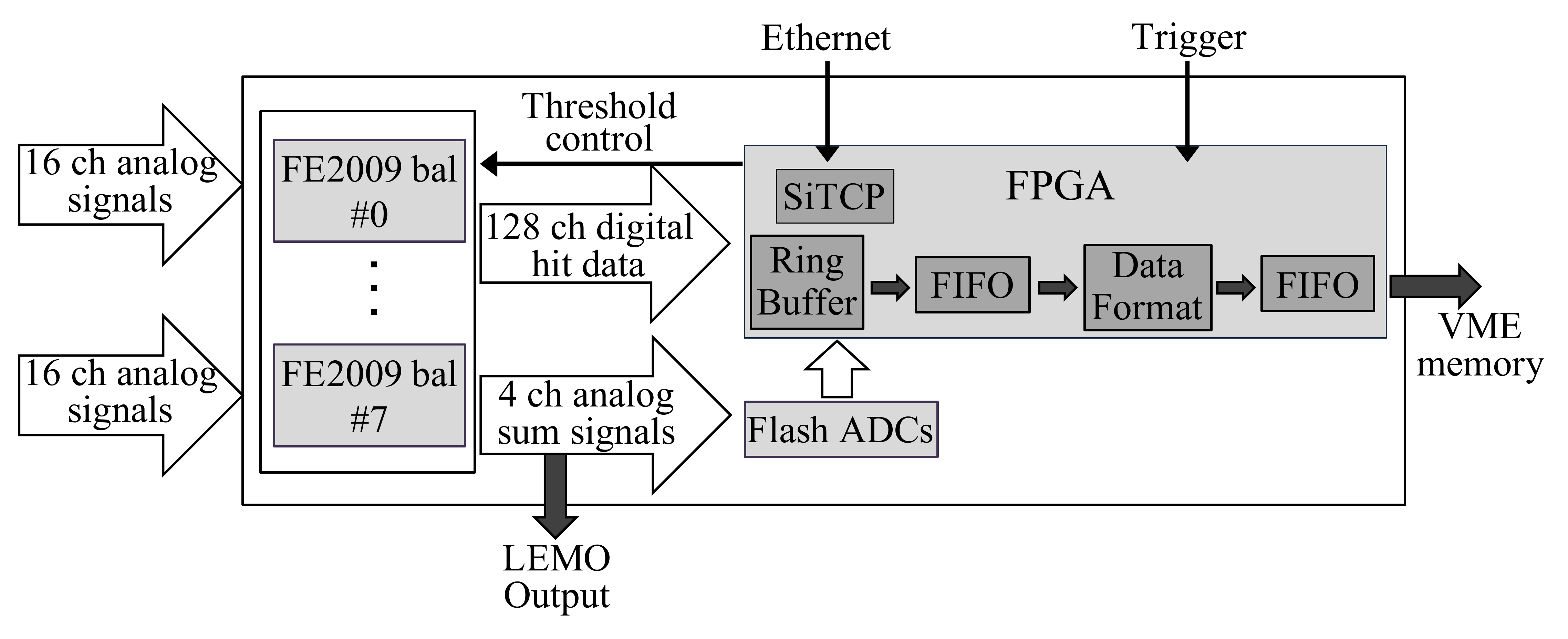}
    \caption{Block diagram of the signal processing in the 
             readout electronics board~\cite{iwaki}.}
    \label{iwaki}
  \end{center}
\end{figure}

Analog signals from the $\mu$-PIC are processed by the readout
electronics boards specially developed by the cosmic-ray group at Kyoto University~\cite{iwaki}.
Each board processes 128 channels.
In MAIKo, four boards are used 
(two for the 256 anode strips and two for the 256 cathode strips).
A block diagram displaying the signal processing in the readout boards 
is shown in Fig.~\ref{iwaki}.

Each readout board consists of eight 
application specific integrated circuit (ASIC) chips 
named FE\-2009bal \cite{iwaki} and a
field programmable gate array (FPGA) chip.
In each ASIC chip, analog signals from 16 adjacent electrodes of the $\mu$-PIC are amplified, 
shaped and discriminated.
The input range of the FE2009bal chips is $\pm$1 pC and the gain of the amplifier is 
0.8 V/pC.
The threshold levels of the 128-channel discriminators are individually
adjusted via the SiTCP communication~\cite{sitcp}.

The discriminated signals of the 128 channels are transferred to the FPGA chip.
The transferred signals are then synchronized with a 100-MHz clock of the FPGA chip.
The timing of the discriminated signals provides the electron drift time,
and the signal duration (time over threshold: TOT) is approximately proportional 
to the charge collected by the $\mu$-PIC.
In this paper, the first clock and the last clock of the TOT is denoted as the leading edge and
the trailing edge, respectively.
The analog outputs from the FE2009bal chips are summed with adjacent 32 strips
and digitized by 8-bit, 25-MHz flash-ADCs (FADC)
before being sent to the FPGA chip.
The summed analog signals are also split and transmitted via
the LEMO connectors on the readout boards
to be used to make a TPC self trigger.
The noise level is typically $\pm$5 mV.
The dynamic range of the FE2009bal chip is $\pm$0.8 V.
During the measurements using the $^{4}$He beam, we adjusted the gas gain of 
the $\mu$-PIC so that the pulse height of beam particles is 30 mV,
which is high enough to discriminate signals from noise.
Under such conditions, the TPC can measure the energy loss of the particles 
up to about 25 times larger than that of the beam particles.
The discriminated 128-ch hit pattern and the 4-ch FADC data are continuously 
stored in a ring buffer in the FPGA chip for 10.24 $\mu$s.
Upon receiving an external trigger signal, the data stored in the ring buffer
is written out to two VME memory modules after the data formatting in the FPGA chip.
The TPC data is acquired on an event-by-event basis 
together with the data from the ancillary detectors
and the beam-line detectors via the VME bus
using the data acquisition (DAQ) program developed at RIBF \cite{BABA201065}.
The trigger signal is usually generated by the Si detectors or 
by discriminating the analog signals from the $\mu$-PIC strips transmitted via the LEMO
connectors on the readout boards.

\begin{figure*}[tbp]
  \begin{center}
    \vspace{-5mm}
    \includegraphics[width=130mm]{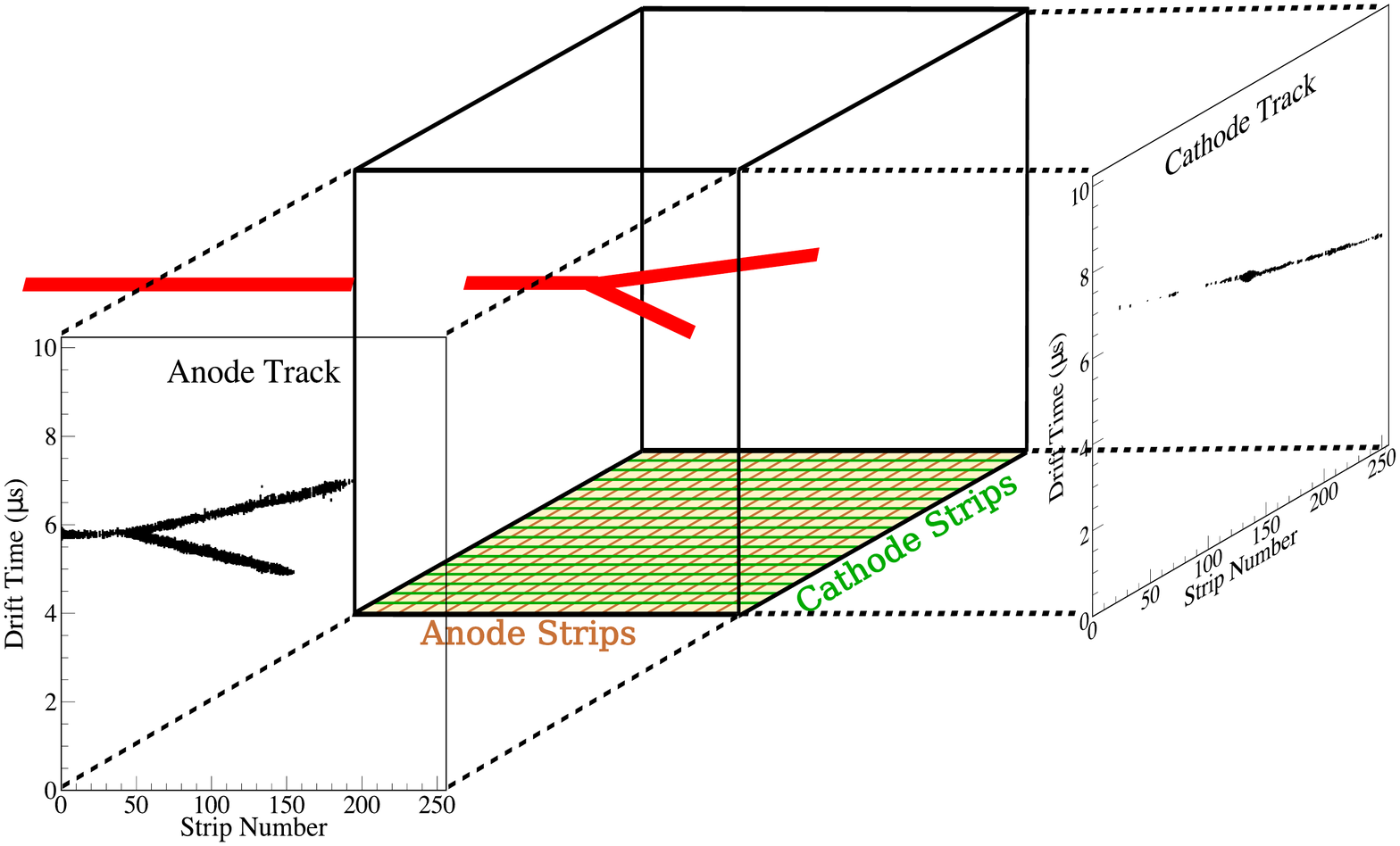}
    \vspace{-10mm}
    \caption{Reconstruction of a track from the TPC data.
      The anode and cathode strip numbers and the electron drift time in each strip
      give the two-dimensional projections of the particle trajectories.
      }
    \label{track_recon}
  \end{center}
\end{figure*}

As shown in Fig. \ref{track_recon},
the anode and cathode strip numbers and the electron drift time in each strip
provide the two-dimensional projections of the trajectories.
The three-dimensional track is reconstructed by combining the
anode and cathode projections.

\subsection{Ancillary detectors}
\label{ancillary}
High-energy recoil particles which punch-through the TPC are detected
with Si detectors followed by CsI(Tl) crystals.
The Si detectors have a sensitive area of 90 $\times$ 60 mm$^{2}$ with a thickness of
325 $\mu$m or 500 $\mu$m depending on experimental requirements.
The CsI(Tl) crystals have a volume of 30 $\times$ 30 $\times$
30 mm$^{3}$.
They are wrapped by enhanced specular reflector (ESR) films from 3M \cite{ESRdata},
and their back side is attached to Si-PIN photodiodes with which
the scintillation photons are detected.

Two Si detectors are installed at the left and the right sides of the TPC
followed by CsI(Tl) detectors.
Four Si detectors are installed at downstream of the TPC.
As illustrated in Fig.~\ref{at_fig},
these four Si detectors are arranged to form a 30-mm square hole 
at the beam position to allow the beam particles to pass through.

The Si and CsI(Tl) detectors are placed close to the TPC field cage
at a distance which is safe enough to avoid the discharge from
the field cage to the Si detectors.

\section{Measurement of the gas gain of the $\mu$-PIC with an alpha source}
\label{alpha}
The MAIKo TPC was commissioned using an $^{241}$Am alpha source 
to measure the gas gain of the $\mu$-PIC.
The TPC chamber was filled with He(93\%)+iso-C$_{4}$H$_{10}$(7\%) 
mixture gas at 430 hPa.
The alpha source was mounted upstream of the TPC.
The distance between the alpha source and the sensitive volume of the TPC was 61 mm.
The range of the 5.48-MeV alpha particles from  $^{241}$Am in the detector gas 
is 264 mm according to the calculation using the SRIM code \cite{ZIEGLER20101818}.
Since this range is long enough for the alpha particles to penetrate the sensitive volume of the TPC,
these alpha particles were detected by one of the Si detectors installed
downstream of the TPC.
Signals from the Si detectors were used to trigger the DAQ system of the TPC.
Measurements were performed for different bias voltages of the grid mesh and the anode
of the $\mu$-PIC.
The electric field between the cathode plate and the grid mesh was kept at 200 V/cm.

\begin{figure}[htbp]
  \begin{center}
    \vspace{-20mm}
    \includegraphics[width=80mm]{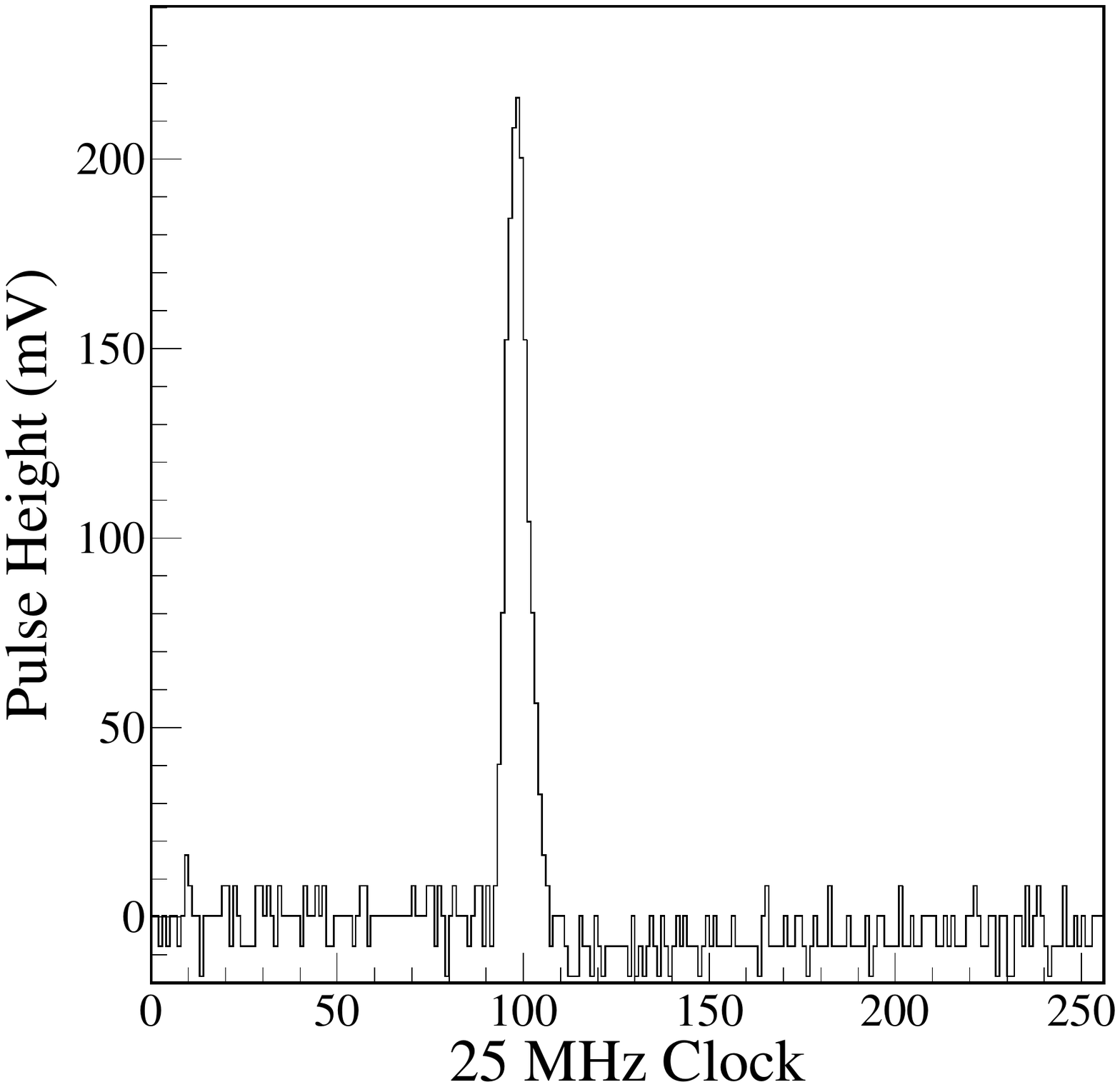}
    \vspace{-20mm}
    \caption{Typical signal acquired with the FADC for an alpha particle emitted from an
      $^{241}$Am alpha source.
      The signal was summed over 32 consecutive anode strips on the $\mu$-PIC
      which have a total width of 12.8 mm.}
    \label{signal}
  \end{center}
\end{figure}

A typical analog signal from 32 anode strips obtained by the FADC is shown in Fig. \ref{signal}.
The flight direction of the alpha particles was almost perpendicular 
to the anode strips.
The charge $Q_{1}$ collected by the anode strips after electron avalanche was derived by
integrating the FADC pulse.
The initial charge $Q_{0}$ caused by the alpha particle is given by
$Q_{0}=e(\Delta E_{\mathrm{He}} / W_{\mathrm{He}} +
 \Delta E_{\mathrm{iC_{4}H_{10}}} / W_{\mathrm{iC_{4}H_{10}}})$.
$\Delta E_{\mathrm{He}}$ and $\Delta E_{\mathrm{iC_{4}H_{10}}}$
are the energy losses of the alpha particles along the passage of 12.8 mm above 
the anode strips in the helium and iso-butane gases at each partial pressure.
$e$ is the charge of an electron.
The $\Delta E_{\mathrm{He}}$ and $\Delta E_{\mathrm{iC_{4}H_{10}}}$ are estimated to be
85 keV and 103 keV by the SRIM code.
$W_{\mathrm{He}}$ (41 eV) and $W_{\mathrm{iC_{4}H_{10}}}$ (26 eV) are
the mean ionization energies of the He and iso-C$_{4}$H$_{10}$
gasses, respectively \cite{pdg}.
The gas gain was determined from $Q_{1}/Q_{0}$.

\begin{figure}[htbp]
  \begin{center}
    \vspace{-5mm}
    \includegraphics[width=75mm]{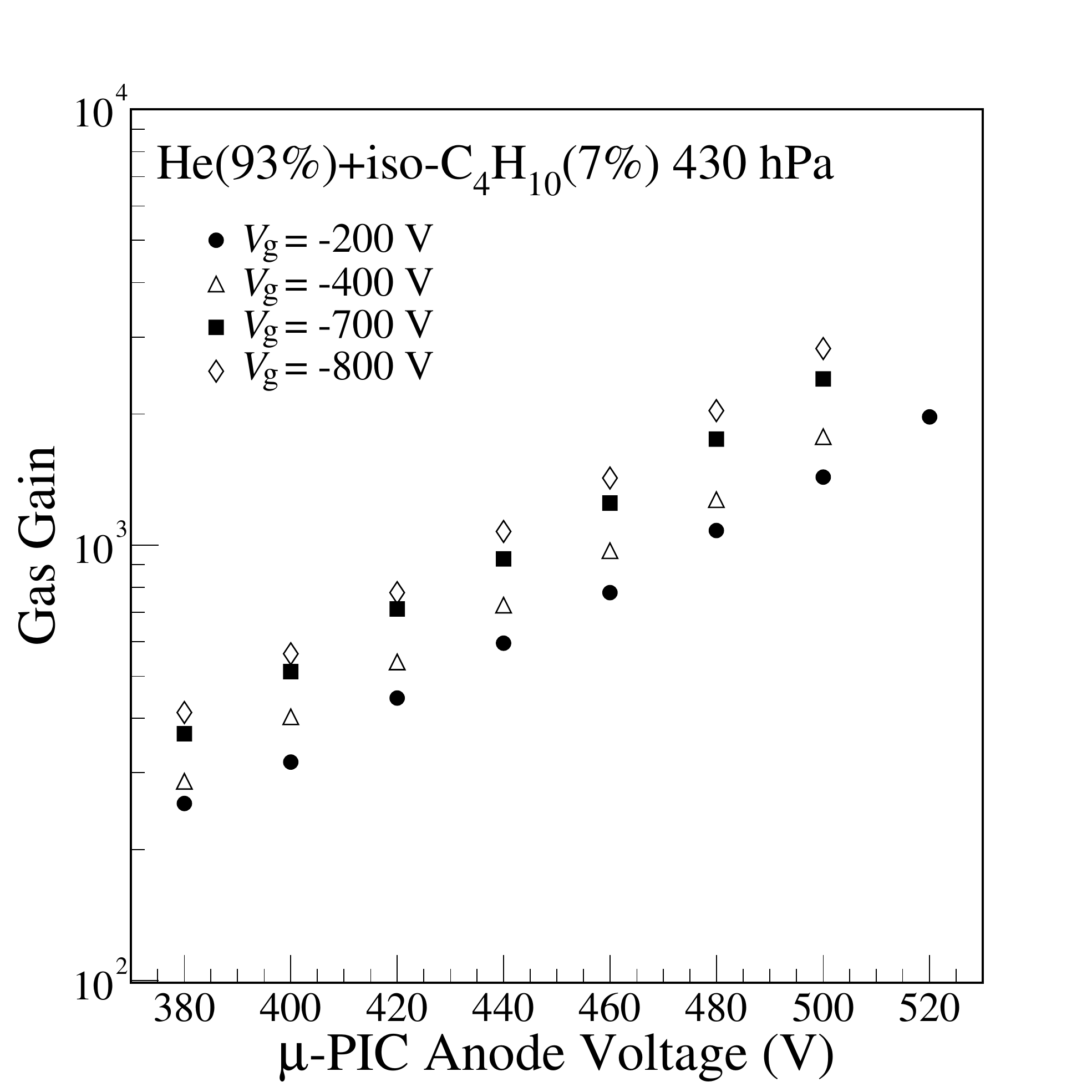}
    \vspace{-1mm}
    \caption{Gas gain of the $\mu$-PIC as a function of the $\mu$-PIC anode voltage.
      The gain was measured with a He(93\%)+iso-C$_{4}$H$_{10}$(7\%) mixture gas at 430 hPa.
      The measurement was performed with different voltages applied to the grid mesh:
      $-200$ V (solid circles), $-400$ V (open triangles), 
      $-700$ V (solid squares), and $-800$ V (open diamonds).}
    \label{gain}
  \end{center}
\end{figure}

Figure \ref{gain} shows the measured gas gains with different voltages applied to the grid mesh
as a function of the bias voltage applied to the anode strips of the $\mu$-PIC.
The gas gain increases with the voltage applied to the grid mesh.
This is because the electric field around the anode of the $\mu$-PIC becomes stronger
with increased grid voltage.
The maximum gain ($\sim$3000) was limited by the discharge of the $\mu$-PIC.

\section{Measurements with a $^{4}$He beam}
\label{beam}
\subsection{Experimental setup}
A test experiment was carried out using a 56-MeV $^{4}$He beam from the azimuthally varying field
(AVF) cyclotron
at RCNP to study the detector performances under high counting rates and to acquire
scattering events.

\begin{figure}[htbp]
  \begin{center}
    \includegraphics[width=80mm]{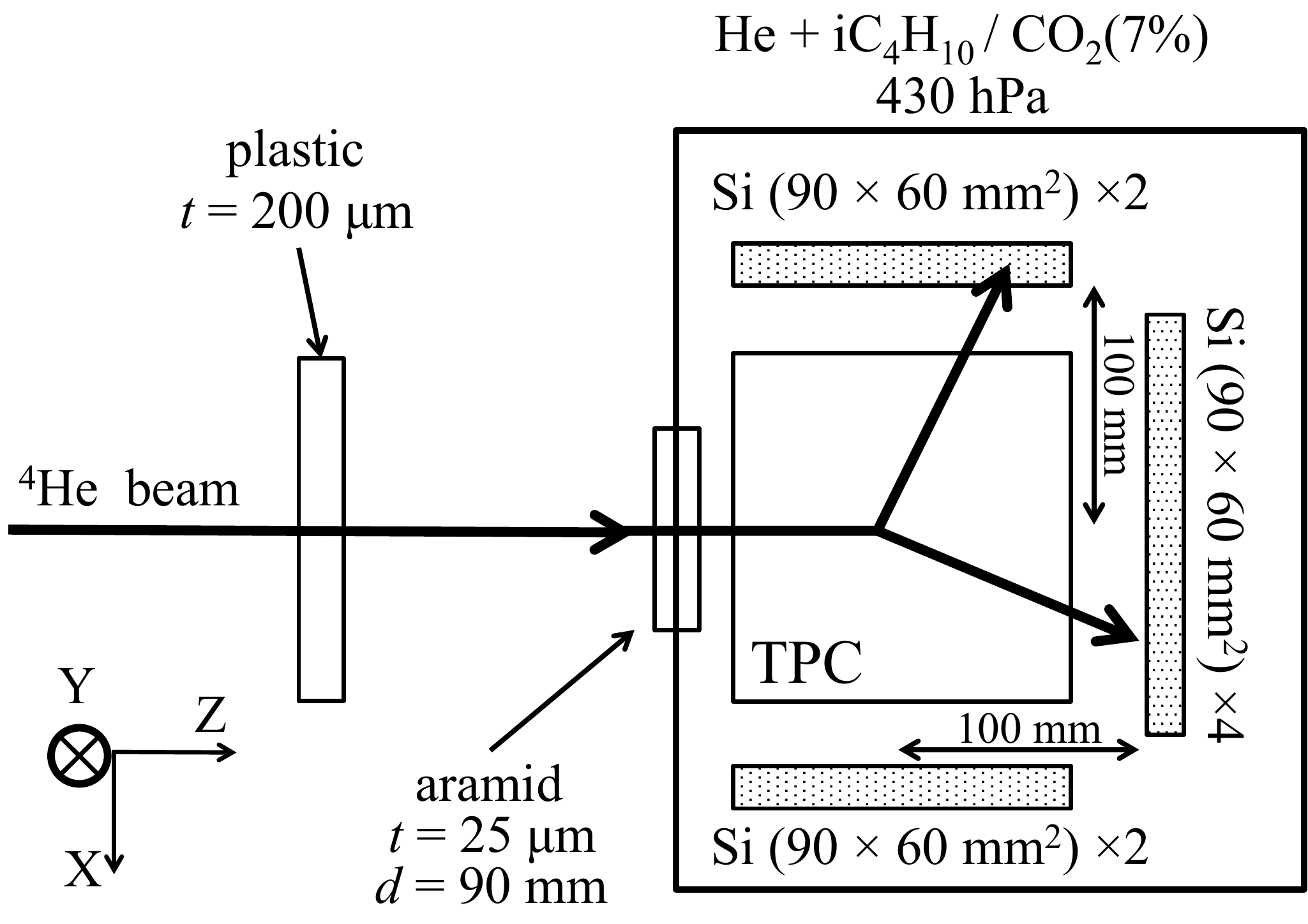}
    \caption{Top view of the experimental setup at the RCNP EN beam line.}
    \label{beam_setup}
  \end{center}
\end{figure}

The experimental setup is shown in Fig.~\ref{beam_setup}.
The MAIKo system was installed at the F2 focal plane of the RCNP EN beam line.
The MAIKo chamber was filled with He(93$\%$) + iso-C$_{4}$H$_{10}$(7$\%$) or
He(93$\%$) + CO$_{2}$(7$\%$) mixture gas at 430 hPa.
The chamber was separated from the vacuum section of the beam line
by a 25-$\mu$m-thick aramid window film.
The diameter of the window film was 90 mm.
The four 500-$\mu$m-thick Si detectors (left and right sides) and 
the four 325-$\mu$m-thick Si detectors (downstream)
were installed 100-mm away from the center of the TPC sensitive volume
as described in Section~\ref{ancillary}.
The four Si detectors at the left and right sides 
were used to create a trigger signal for the scattering events.
A plastic scintillator with a thickness of 200 $\mu$m was installed upstream of the
TPC to measure the beam intensity.
This scintillator was also used to define the start timing of the TPC.
The energy losses of the $^{4}$He beam by the plastic scintillator and the
aramid foil are 3.0 MeV and 0.4 MeV, respectively.
The energy and angular stragglings in the two materials are
0.1 MeV and 0.3 deg, respectively, 
which are negligible in the present analysis.

\subsection{Angular resolution}
\label{secangreso}
The angular resolution of the TPC was evaluated from the $^{4}$He beam events.
The data was taken with both He + iso-C$_{4}$H$_{10}$(7$\%$)
and He + CO$_{2}$(7$\%$) gases at 430 hPa.
The voltage parameters of the TPC (cathode plate: $V_{\mathrm{c}}$,
grid mesh: $V_{\mathrm{g}}$, 
and $\mu$-PIC anode: $V_{\mathrm{a}}$, see Fig.~\ref{cage}) for both of the gas mixtures
are listed together with the drift velocity $v$ in Table \ref{vpara}.
These operating voltages were chosen so as to optimize the
gas gain without discharge.
The drift velocities were calculated with the Magboltz code \cite{mag}.
We confirmed that the calculated drift velocities agreed with 
the measured drift velocities within 5\%.

\begin{table}[htb]
  \begin{center}
    \caption{High voltages applied to the cathode plate
      ($V_{\mathrm{c}}$), the grid mesh ($V_{\mathrm{g}}$), and
      the anode of $\mu$-PIC ($V_{\mathrm{a}}$) during the commission with a $^{4}$He beam.
      The drift velocities ($v$) with these high voltages estimated by the Magboltz simulation
      are also listed.}
    \vspace{2mm}
    \begin{tabular}{c c c c c} \hline \hline
      Gas & $V_{\mathrm{c}}$ (V) & $V_{\mathrm{g}}$ (V) &
      $V_{\mathrm{a}}$ (V) & $v$ (cm/$\mu$s)\\ \hline
      He+iC$_{4}$H$_{10}$ & $-3500$ & $-700$ & 500 & 1.59   \\ 
      He+CO$_{2}$         & $-3200$ & $-980$ & 550 & 1.34   \\ \hline \hline
    \end{tabular}
    \label{vpara}
  \end{center}
\end {table}


\begin{figure}[h]
  \begin{center}
    \vspace{-2.0mm}
    \includegraphics[width=70mm]{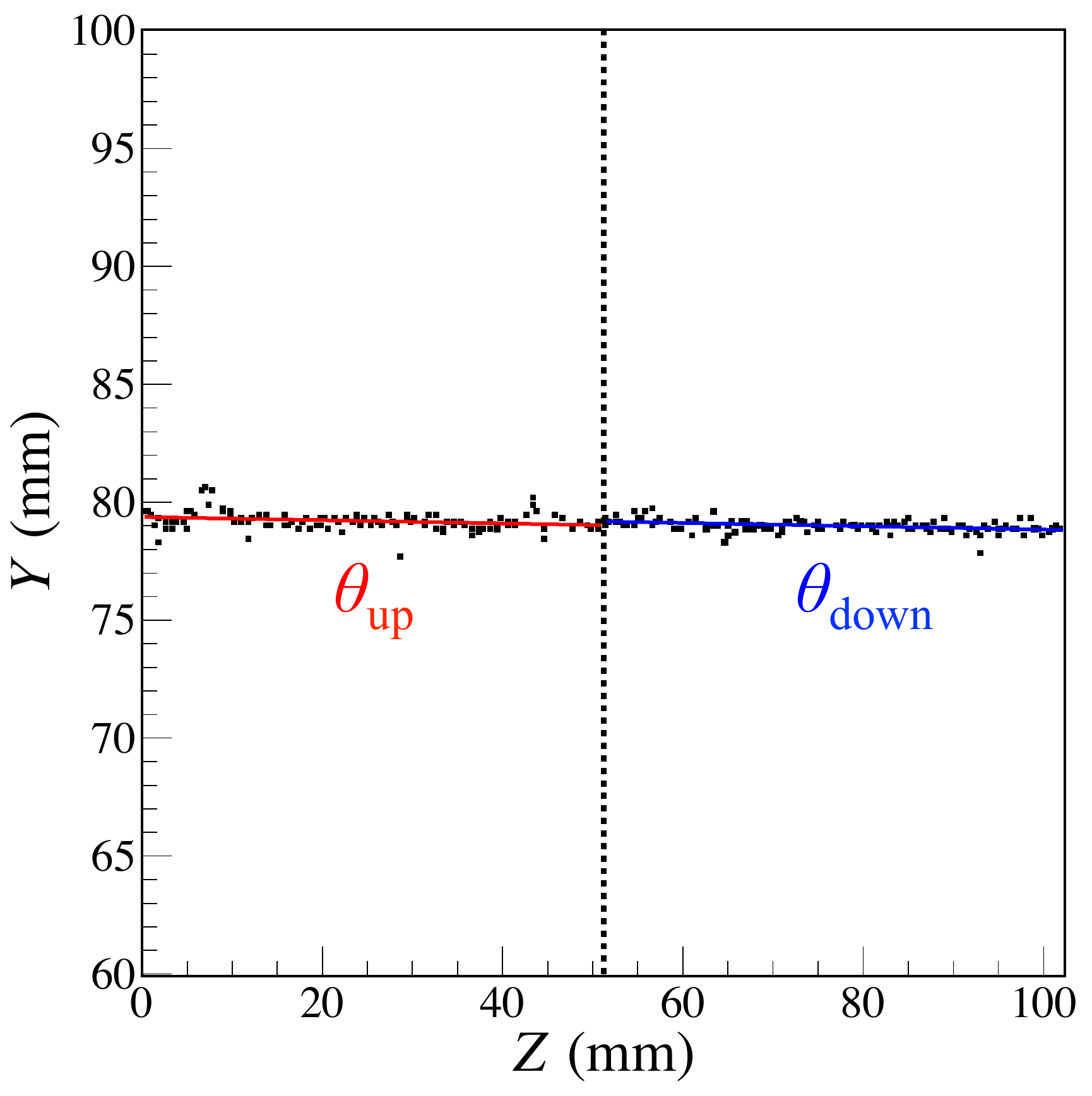}
    \caption{Example of a $^{4}$He beam track of the anode strips.
             The anode strips were perpendicular to the beam axis.
             The $Z$ coordinate was determined by the strip number multiplied by the 
             strip pitch of 400 $\mu$m, and the $Y$ coordinate was determined by the 
             electron drift time multiplied by the drift velocity.
             Only the middle clocks between the leading edge and the trailing edge 
             are shown.
             The track was separated into upstream and downstream regions and each
             track was fitted to a straight line (solid line) to determine the angle
             of the beam.}
    \label{afit}
  \end{center}
\end{figure}

\begin{figure}[h]
  \begin{center}
    \vspace{-1mm}
    \includegraphics[width=70mm]{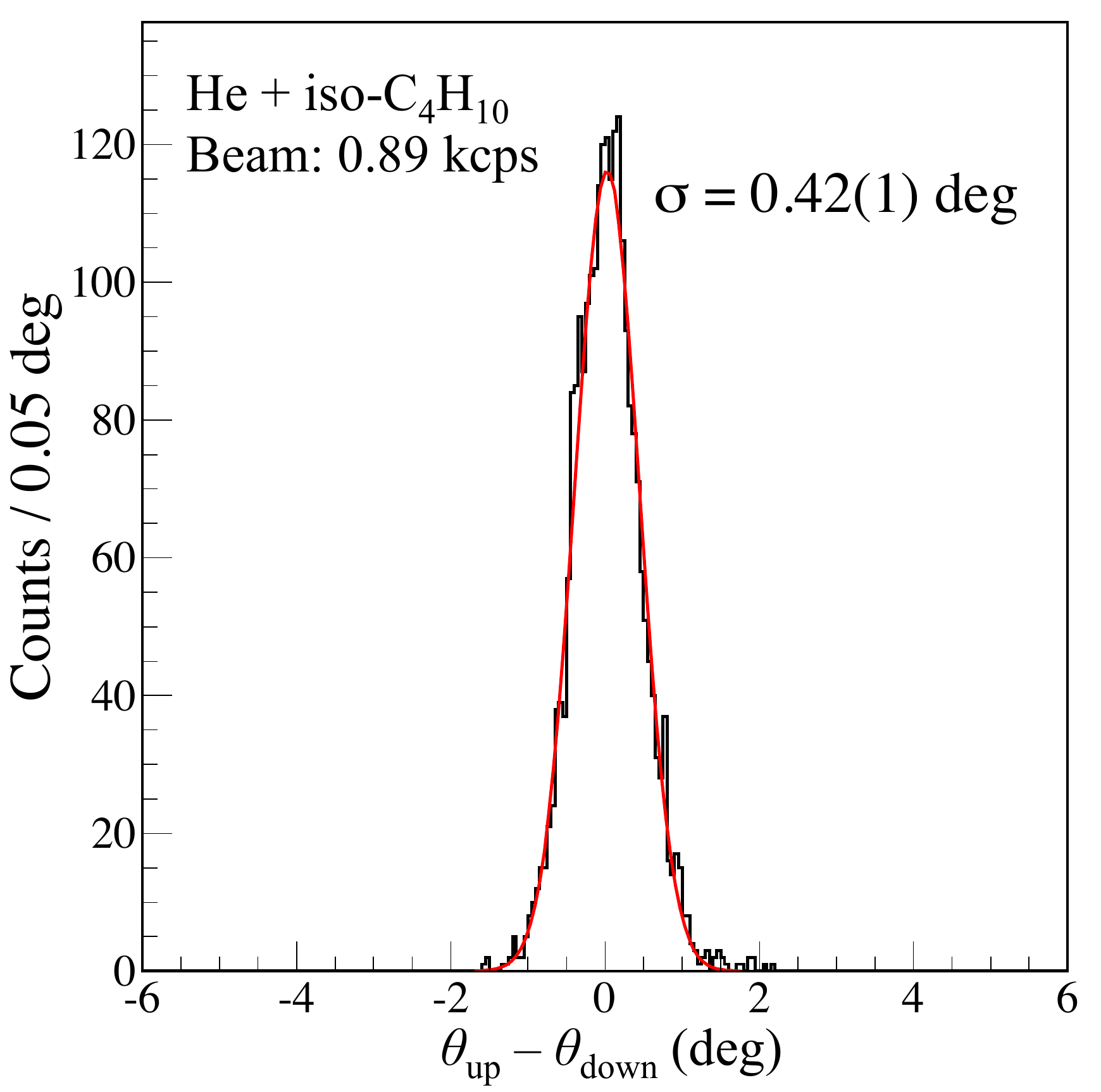}
    \caption{The distribution of the difference of the beam angle between the upstream and
             the downstream regions.
             Each angle was calculated by the fitting the
             tracks to a straight line.
             The histogram was fitted by a Gaussian function (red solid line) to determine
             the angular resolution.}
    \label{angdiff}
  \end{center}
\end{figure}

Figure \ref{afit} shows a typical track of a beam particle reconstructed from the anode data.
The anode strips were perpendicular to the beam axis.
The $Z$ coordinate was determined from the anode strip number, and the $Y$ coordinate was determined
from the drift time multiplied by the drift velocity.
As described in Section \ref{readout}, the time over threshold was recorded for each strip, but
only the middle clock between the leading edge and the trailing edge was
used in the analysis.

To evaluate the angular resolution, the tracks of the beam particles were divided into
two parts, upstream and downstream.
Each part contains data for the respective 128 strips (51.2 mm).
These tracks were fitted separately by straight lines to obtain the angles of the track
($\theta_{\mathrm{up}}$ and $\theta_{\mathrm{down}}$).
$\theta_{\mathrm{up}}$ should agree with $\theta_{\mathrm{down}}$ within the angular straggling
of the $^{4}$He beam (less than 0.05$^{\circ}$).
However, due to the limited angular resolution of the TPC, the distribution of
$\theta_{\mathrm{up}} - \theta_{\mathrm{down}}$ is smeared out from 0$^{\circ}$ 
as shown in Fig. \ref{angdiff}.
The standard deviation ($\sigma_{\mathrm{up-down}}$) of the distribution was determined
by fitting a Gaussian function (the red line in the figure).
Assuming the angular resolution for the upstream and the downstream regions to be the same,
the angular resolution for the beam particles was calculated as
$\sigma_{\mathrm{beam}}=\sigma_{\mathrm{up-down}} / \sqrt{2}$.

\begin{figure}[hbtp]
  \begin{center}
    \vspace{-7mm}
    \includegraphics[width=75mm]{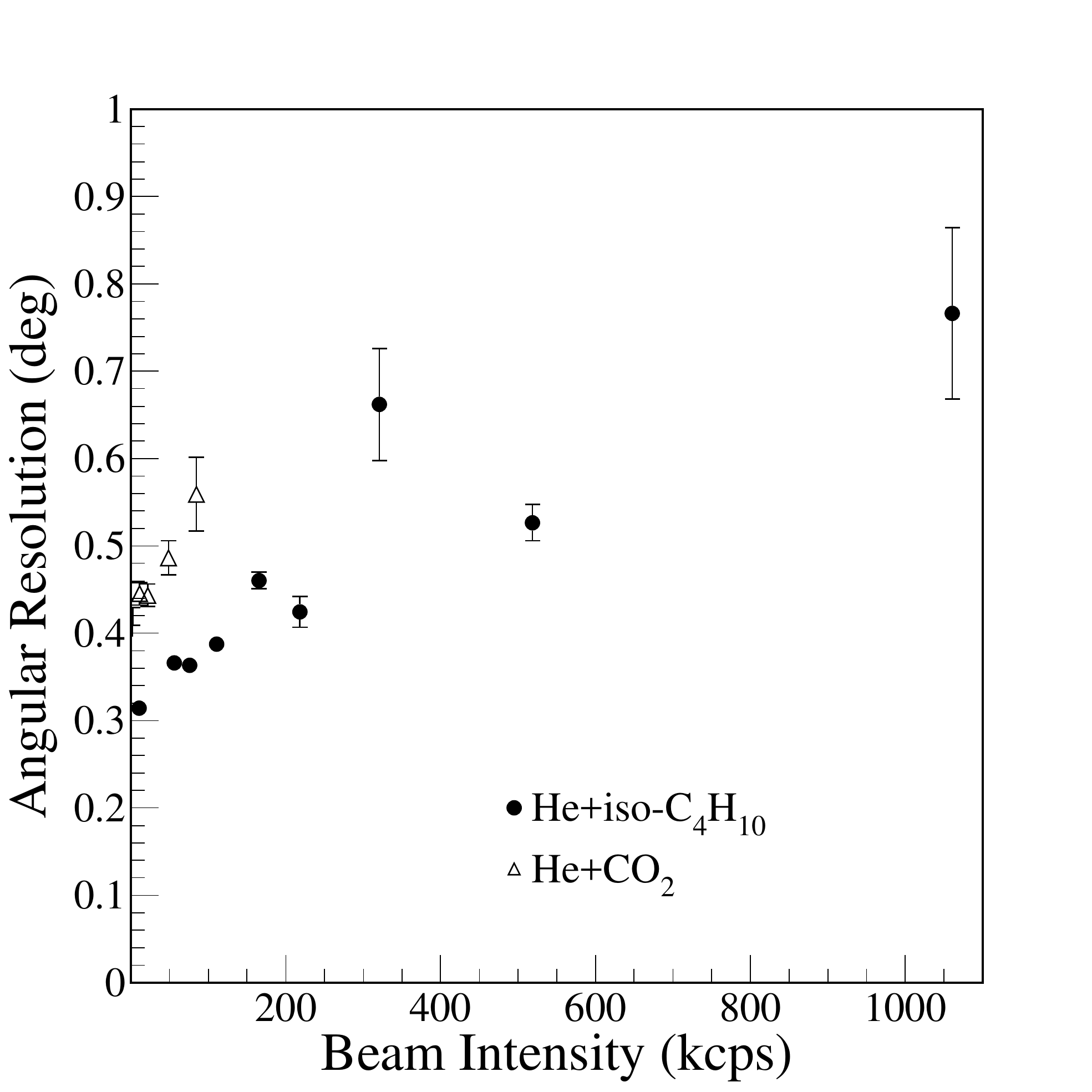}
    \vspace{-0mm}
    \caption{Angular resolution of the TPC for the beam particles ($^{4}$He at 53 MeV) 
             as a function of the beam intensity.
             The solid circles represent the resolution with the 
             He(93\%)+iso-C$_{4}$H$_{10}$(7\%)
             gas mixture and the open triangles represent the resolution with the 
             He(93\%)+CO$_{2}$(7\%) gas mixture.
             The pressure of the gasses was 430 hPa.
             The beam intensities were increased until the discharge occurred 
             at the $\mu$-PIC.}
    \label{ang_reso}
  \end{center}
\end{figure}

Figure \ref{ang_reso} shows the angular resolution measured at various $^{4}$He beam
intensities.
Whereas operation with the He+CO$_{2}$ gas was unstable
at beam intensities higher than 100 kcps due to the discharge of the $\mu$-PIC,
stable operation at beam intensities higher than 1000 kcps was possible with the
He+iso-C$_{4}$H$_{10}$ gas.
We note that the high voltage applied
to the anode of the $\mu$-PIC was increased to 550 V 
during the measurements with beam intensities of 519 and 1061 kcps
to keep the pulse height for the
beam particles same as that for the lower beam intensity.
The angular resolution worsened with increasing beam intensity
because of the buildup of space charge inside the field cage.


\section{Scattering events}
\label{scattering}
\subsection{Examples of tracks}
Scattering events of the $^{4}$He beam off the TPC gas were acquired to
develop an event reconstruction algorithm.
The TPC was operated with the He(93\%)+iso-C$_{4}$H$_{10}$(7\%) gas at 430 hPa.
The intensity of the $^{4}$He beam was 50 kcps.
The trigger signals for the scattering events were generated by the Si detectors
installed at left and right sides of the TPC.

\begin{figure*}[htbp]
  \begin{center}
    \vspace{-10mm}
    \subfigure[Anode track.]{\includegraphics[width=70mm]{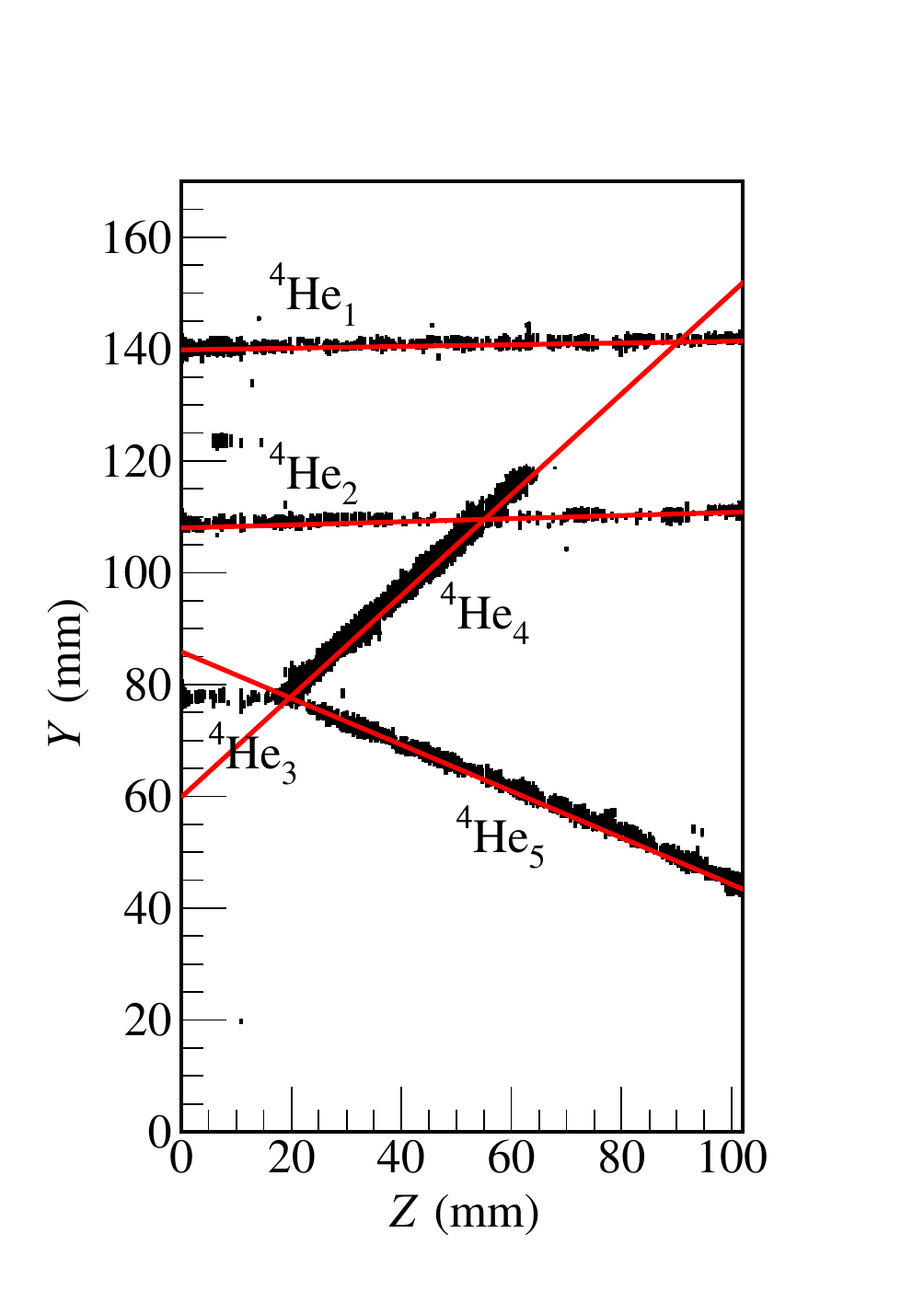}
      \label{scaa}}
    \subfigure[Cathode track.]{\includegraphics[width=70mm]{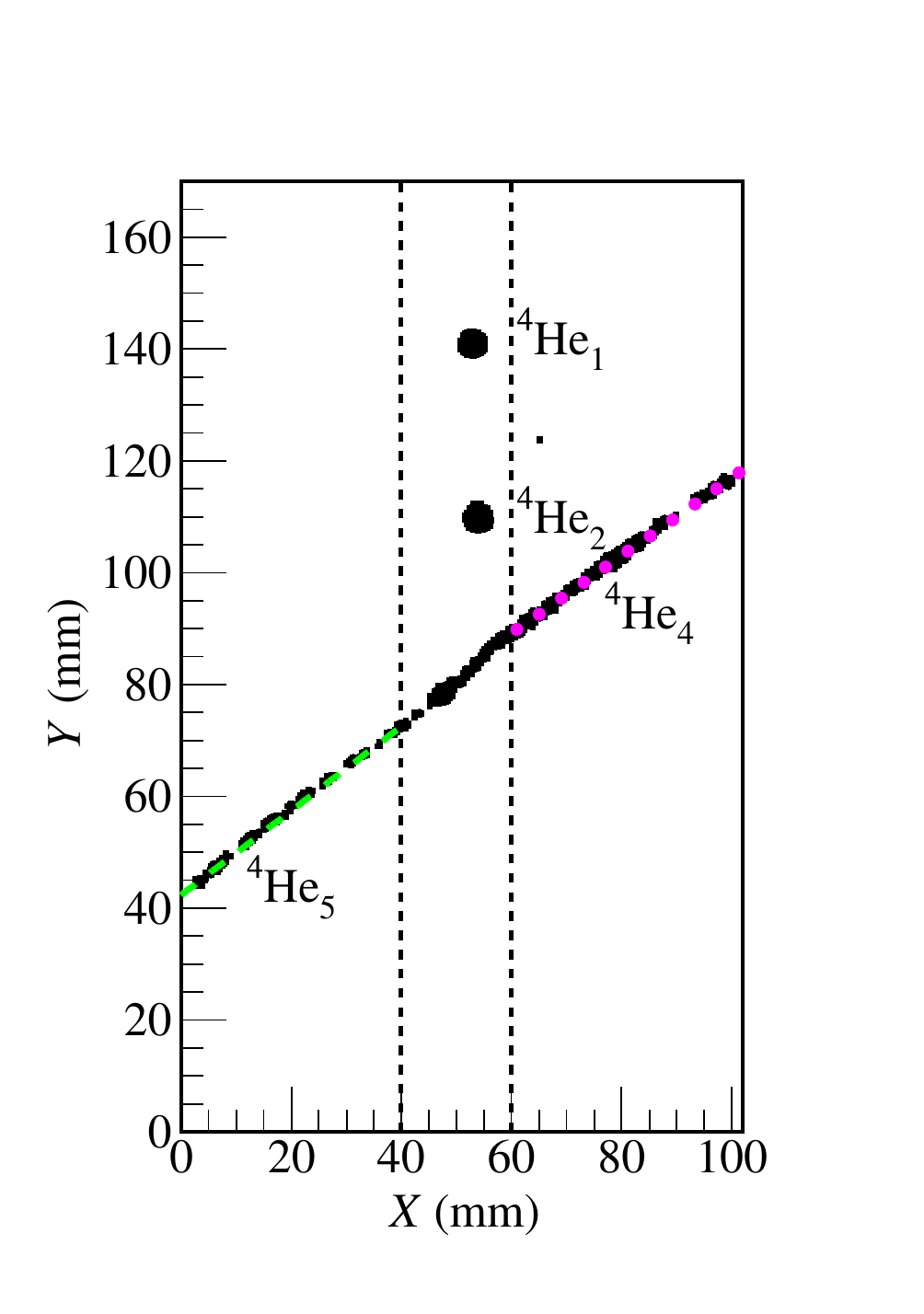}
      \label{scac}}
    \caption{Track sample of a $^{4}$He+$^{4}$He scattering.
             The $Z$ position in the left panel and the $X$ position in the 
             right panel were determined from the $\mu$-PIC strip number.
             The $Y$ position in both panels were determined from the  
             electron drift time multiplied by the electron drift velocity.
             The red-solid lines, green-dashed line, and magenta-dotted line
             represent the tracks identified by the Hough 
             transform algorithm described in the text.
             The vertical dashed lines are drawn at $X=$ 40 and 60 mm.}
    \label{sca_track}
  \end{center}
\end{figure*}

\begin{figure*}[htbp]
  \begin{center}
    \vspace{-10mm}
    \subfigure[Anode track.]{\includegraphics[width=70mm]{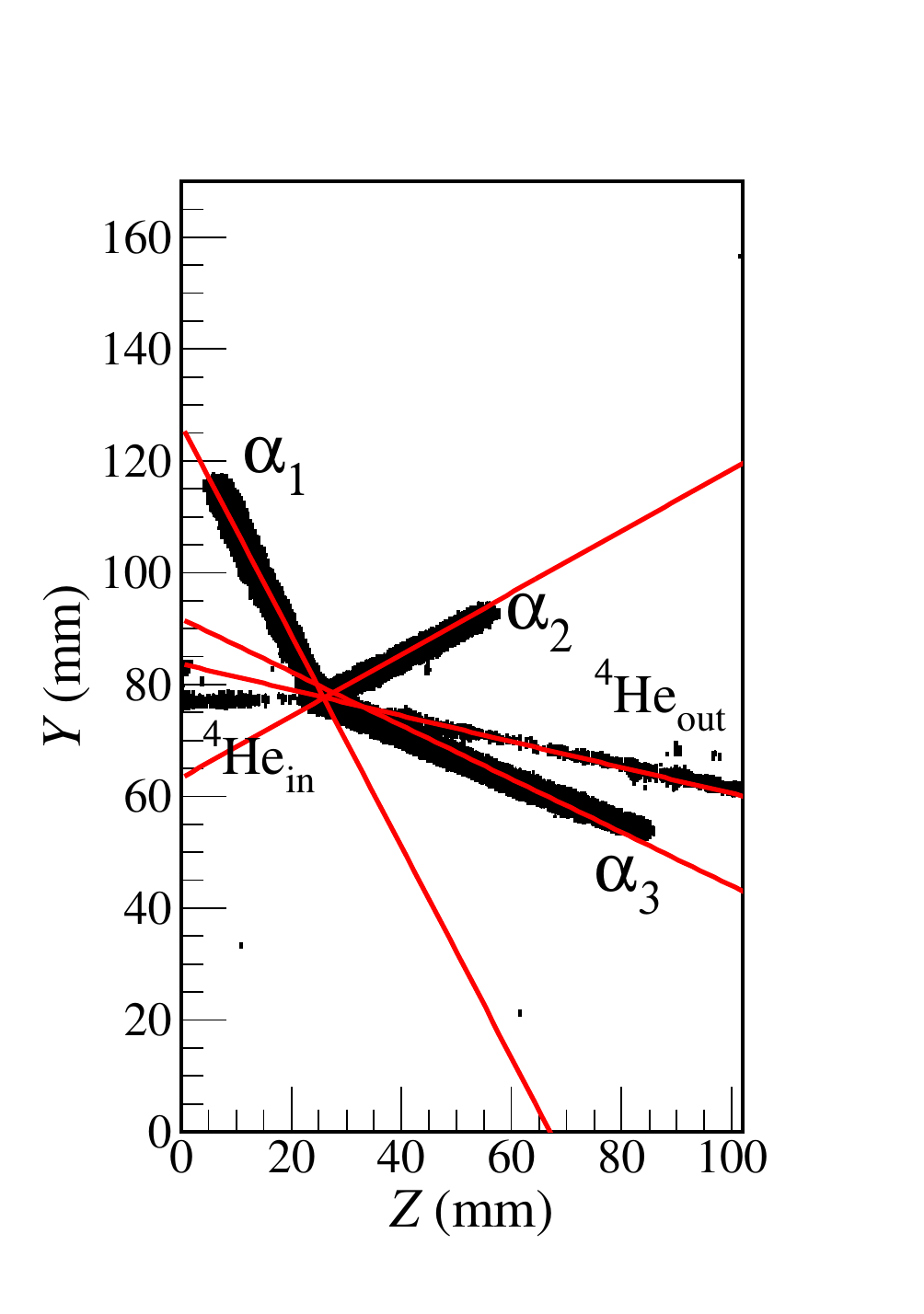}
      \label{4bodya}}
    \subfigure[Cathode track.]{\includegraphics[width=70mm]{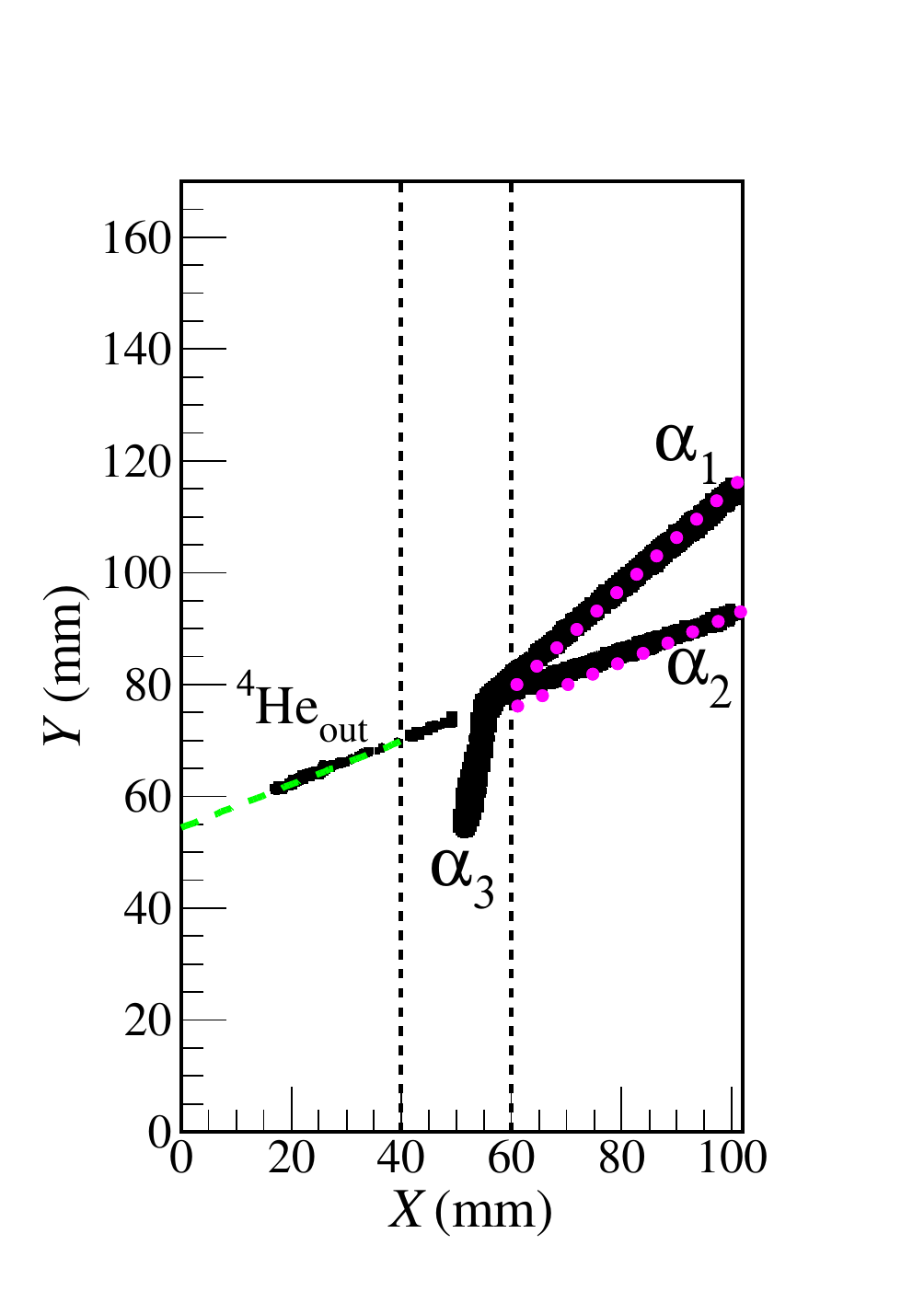}
      \label{4body}}
    \caption{Similar to Fig. \ref{sca_track}, but for 
             a $^{4}\mathrm{He}+^{12}\mathrm{C}$ inelastic scattering.
             The $^{12}$C nucleus contained in the iso-C$_{4}$H$_{10}$ gas decayed 
             into three alpha particles ($\alpha_{1}$, $\alpha_{2}$, and $\alpha_{3}$)
             immediately after being excited to above the alpha decay threshold.}
    \label{4body_track}
  \end{center}
\end{figure*}

The acquired scattering data contains not only
the $^{4}$He+$^{4}$He elastic scattering events,
but also scattering from $^{1}$H or $^{12}$C in the iso-C$_{4}$H$_{10}$ gas.
Typical measured tracks are presented
in Figs. \ref{sca_track} and \ref{4body_track}.
The $X$ and $Z$ coordinates were determined from the cathode and anode strip numbers on the $\mu$-PIC,
and the $Y$ coordinate
was determined from the electron drift time multiplied by the drift velocity.

Figure \ref{sca_track} shows the tracks of a $^{4}$He+$^{4}$He elastic scattering event.
The incident $^{4}$He beam (labeled as $^{4}$He$_{3}$ in the figure) was scattered from
the $^{4}$He gas inside the TPC sensitive volume.
From the vertex point, the recoil and the scattered $^{4}$He particles
(denoted by $^{4}$He$_{4}$ and $^{4}$He$_{5}$, respectively)
were emitted and escaped from the TPC.
The tracks of the unreacted pile-up beam particles
(denoted as $^{4}$He$_{1}$ and $^{4}$He$_{2}$, respectively)
were also recorded at the same time.
The particles $^{4}$He$_{1}$ and $^{4}$He$_{2}$
did not actually pass at $Y\sim$80 mm and $Y\sim$140 mm, respectively.
The heights of the tracks of these particles were wrongly determined
because their arrival timings were different from that of $^{4}$He$_{3}$.
In addition to the $^{4}$He tracks, small points, which do not belong to the particle
trajectories, were also recorded.
These signals could be attributed to the noise in the circuit or X rays from the gas atoms.

Figure \ref{4body_track} shows the tracks of a 
$^{4}\mathrm{He}+^{12}\mathrm{C}$ inelastic scattering event.
The $^4$He beam ($^{4}$He$_{\mathrm{in}}$) was scattered from $^{12}$C
in the iso-C$_{4}$H$_{10}$
gas and exited the TPC ($^{4}$He$_{\mathrm{out}}$).
The $^{12}$C target nucleus was excited to above the alpha decay threshold
at $E_x=7.27$~MeV, and as a result, three alpha particles 
($\alpha_{1}$, $\alpha_{2}$ and $\alpha_{3}$) were emitted.
While $\alpha_{1}$ and $\alpha_{2}$ escaped from the TPC and were detected
by the Si detectors, $\alpha_{3}$ stopped inside the TPC.
Because these decay alpha particles lost more energies 
than $^{4}$He$_{\mathrm{in}}$ and
$^{4}$He$_{\mathrm{out}}$, the induced pulse heights were higher.
Thus the TOTs became longer, and the observed tracks of the decayed particles 
were thicker.

As seen in Figs. \ref{sca_track} and \ref{4body_track}, many trajectories of incident particles,
scattered particles, decay particles, unreacted particles, {\it etc.} were recorded in the
scattering events.
For the reconstruction of the scattering events, each trajectory in the anode and cathode
images must be extracted separately.
Here, we adopted an image processing algorithm, namely the Hough transform method.

\subsection{Track finding algorithm}
The Hough transform method was originally invented to analyze the bubble chamber pictures \cite{hough}.
It allows the extraction of features such as lines and circles in a picture without
knowing how many features are contained.
This algorithm is now widely applied to image recognition 
because of its high immunity to random noise in images compared to other methods
like the least-square technique.

\begin{figure}[htbp]
  \vspace{-3mm}
  \begin{center}
    \includegraphics[width=90mm]{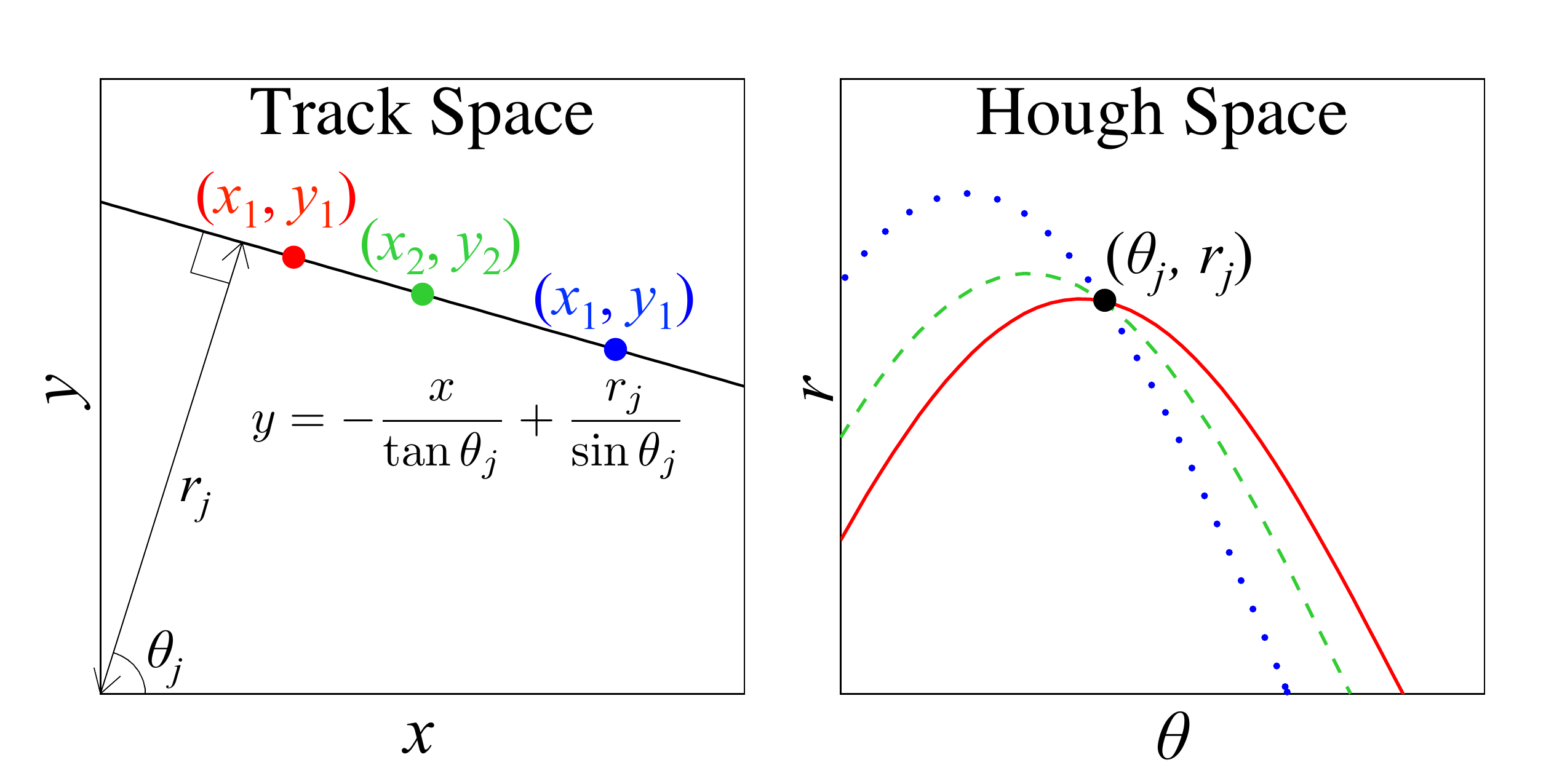}
    \caption{Example of the Hough transformation.
             Three points at $(x_{1},y_{1})$, $(x_{2},y_{2})$, and $(x_{3},y_{3})$ in the
             left panel are transformed into the red-solid, green-dashed,
             and blue-dotted curves, respectively, according to 
             Eq.~(\ref{hougheq}).
             The intersection point of the curves $(\theta_{j}, r_{j})$ gives the
             equation of the line in the track space.}
    \label{hough_exp}
  \end{center}
\end{figure}

The anode and cathode track images obtained by MAIKo contain 256-strip $\times$ 1024-clock pixels.
The coordinates of the $i$-th pixel are expressed as $(x_{i}, y_{i})$ in the track space,
as illustrated in the left panel of Fig.~\ref{hough_exp}.
In the Hough transformation, a hit pixel at $(x_{i}, y_{i})$ is
transformed into a curved line in the $(\theta,r)$ parameter space (Hough space) 
according to the following formula \cite{hough2}:

\begin{equation}
  r=x_{i} \cos \theta + y_{i} \sin \theta .
\label{hougheq}
\end{equation}
For example, the points at $(x_{1},y_{1})$, $(x_{2},y_{2})$, and $(x_{3},y_{3})$ in the track space
are transformed into the red-solid, green-dashed, and blue-dotted curves in the Hough space, respectively.
A point at $(\theta_{j}, r_{j})$ in the Hough space specify one straight line in the track space
as given by the following formula.
\begin{equation}
  y=-\frac{x}{\tan \theta _{j}} + \frac{r_{j}}{\sin \theta _{j}} .
\label{trackeq}
\end{equation}
As shown in the left panel of Fig.~\ref{hough_exp}, 
$r_{j}$ corresponds to the distance between the straight line and the origin.
$\theta_{j}$ corresponds to the angle between the $x$-axis and
the perpendicular line from the origin.

If the pixels in the anode or cathode track space lie on a straight line,
their transformed curves intersect at one point at $(\theta_{j}, r_{j})$ in the Hough space.
Thus, the coordinates of the intersection point give the particle track according to Eq. (\ref{trackeq}).

\begin{figure}[htbp]
  \begin{center}
    \vspace{2mm}
    \includegraphics[width=80mm]{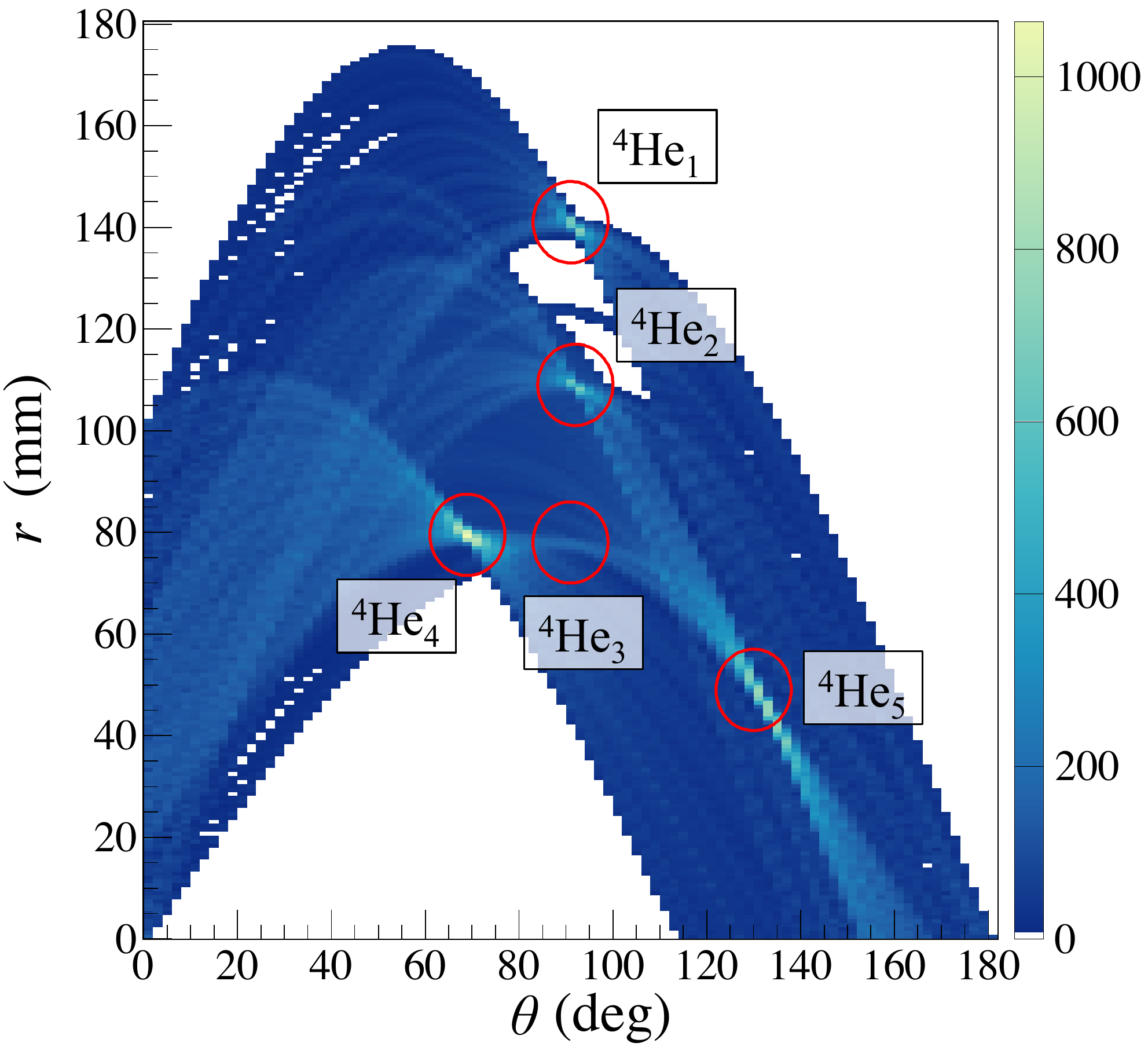}
    \vspace{1.0mm}
    \caption{Two-dimensional histogram in the Hough space made from
             the anode image shown in Fig.\ref{scaa}.
             The circled peaks correspond to the particle trajectories
             in Fig. \ref{scaa}.}
    \label{hough}
  \end{center}
\end{figure}

In the present analysis, the transformed curves were discretized into pixels
and these pixels were booked into a two-dimensional histogram in the Hough space.
For example, the anode image in Fig. \ref{scaa} was transformed into the Hough histogram
shown in Fig. \ref{hough}.
The bin size of the histogram was set to 2$^{\circ} \times 0.78$ mm. 
The tracks with sizable length and width made clear peaks
as labeled and circled in the Hough histogram, 
whereas the small spots caused by noise did not.
By finding the pixel with the maximum count in the Hough histogram, which is easy to implement in a 
computer program, the longest track in the anode image was determined. 

After the first track was extracted, pixels along the first track were eliminated
from the image, and the remaining pixels were again transformed to the Hough space.
The maximum pixel in the second Hough histogram provided the second track in the image.
The above procedures were repeated until the number of content in the maximum pixel
became lower than a threshold (150 counts).
In this way, multiple tracks were separately determined.

The tracks extracted from the anode image in Fig. \ref{scaa}
are shown as the red-solid lines in the figures.
In the cathode track finding, the image was divided into two regions:
$X<40$ mm and $X>60$ mm in order to avoid the beam tracks
because the beam tracks were observed as the circular spots around $X=50$ mm in the
cathode image as seen in Fig.~\ref{scac}.
The cathode strips at $40 \ \mathrm{mm} < X < 60 \ \mathrm{mm}$ 
were not taken into account in the track finding.
The tracks in the $X<40$ mm and  $X>60$ mm regions were independently extracted
as indicated by the green-dashed line and the magenta-dotted lines, respectively.
The track of $\alpha_{3}$ was not extracted from the cathode image because
this track remained within 40 mm $<X<$ 60 mm.

\subsection{Analysis of the $^{4}$He+$^{4}$He elastic scattering}
Because the relative angle between the scattered and the recoil particles is
always 90$^{\circ}$ in the elastic scattering of identical particles,
this reaction is a suitable benchmark to evaluate the angular resolution
of the detector.

After the track finding in the anode and cathode images by the
Hough transform algorithm was finished, the following two conditions were imposed to
identify candidates for elastic scattering events.
First, two tracks with tilted angles more than 4$^{\circ}$ from the $Z$-axis
were found in the anode image.
Second, a single track was found in both of the $X<40$ mm and $X<60$ mm regions
of the cathode image.
The tracks in Fig.~\ref{sca_track} satisfy the above conditions, whereas
the tracks in Fig.~\ref{4body_track} do not because three
tilted tracks were observed in the anode image and
the two tracks were observed in the $X>60$ mm region 
of the cathode image (magenta-dotted lines).

\begin{figure}[htbp]
  \begin{center}
    \vspace{-7mm}
    \includegraphics[width=80mm]{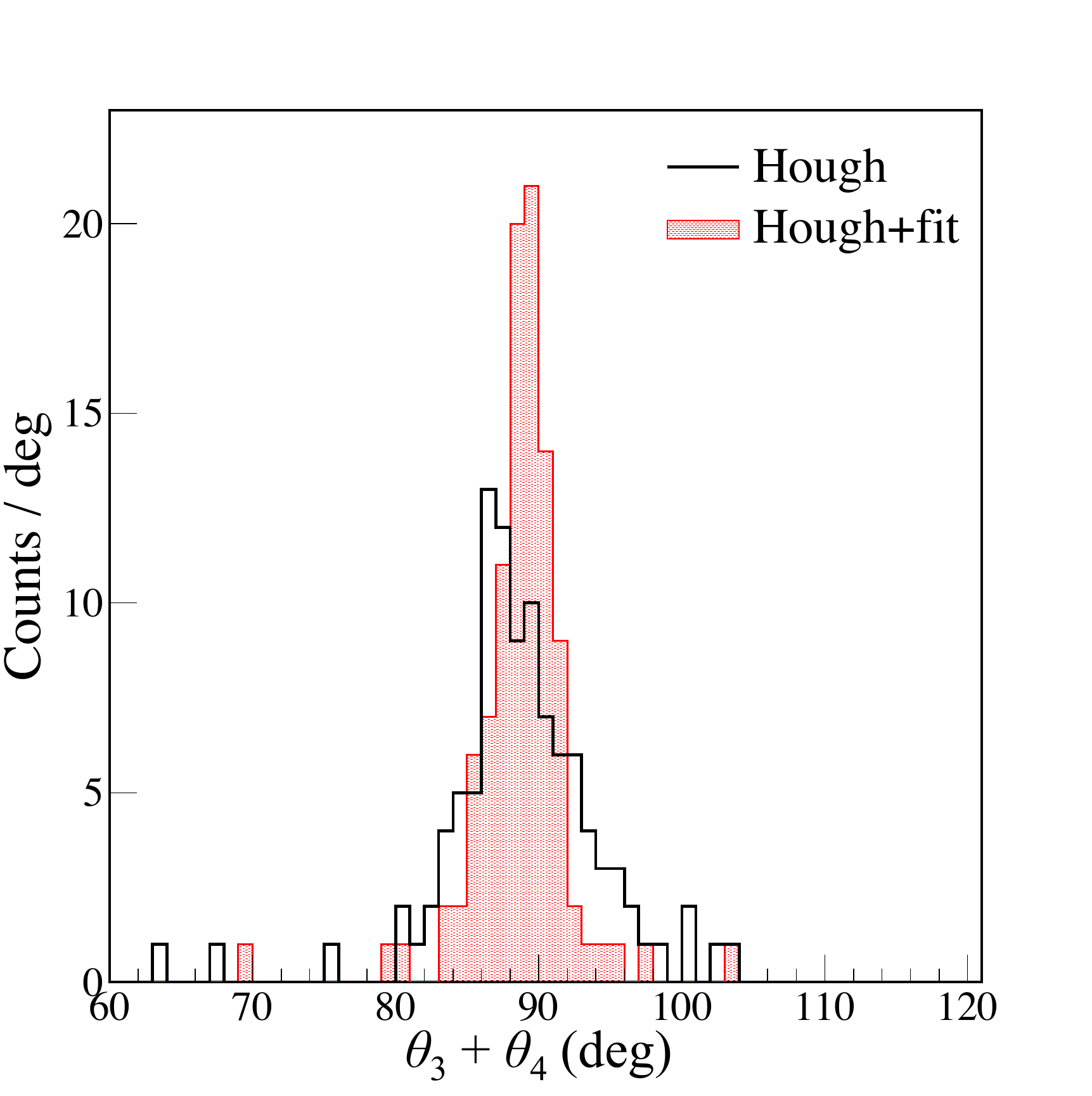}
    \caption{Distribution of the relative angle between the two emitted particles in the $^{4}$He+$^{4}$He elastic scattering.
             The black open histogram represents the angle determined directly from the
             Hough transformation [$\sigma(\theta_{3}+\theta_{4})=3.9^{\circ}$].
             The angle obtained by the fitting after the Hough transformation is shown
             as the red hatched histogram [$\sigma(\theta_{3}+\theta_{4})=1.9^{\circ}$].}
    \label{rela_ang}
  \end{center}
\end{figure}

The three dimensional trajectories of the two particles
observed in the candidates for the elastic scattering events
were reconstructed by combining the
two tracks in the anode and cathode images,
and the opening angle between the two particles was determined.
The reconstructed opening angle between the two trajectories ($\theta_3+\theta_4$)
is shown in Fig.~\ref{rela_ang}.
The black line shows the histogram of the opening angle
calculated directly from the discretized $\theta$ values 
of the peaks in the Hough histograms.
The distribution shows a clear peak at 90$^{\circ}$,
and this result suggests that the elastic scattering events
were successfully identified in the present analysis.

The standard deviation of the distribution [$\sigma(\theta_{3}+\theta_{4})$]
is 3.9$^{\circ}$.
This angular resolution can be improved by further analysis,
because the accuracy of the present analysis is limited by 
the granularity of the Hough histogram.
In the present analysis, the angular bin size of the Hough histogram was 2$^{\circ}$.
With smaller bin size, we can expect to improve the angular resolution.
However, the analysis with a finer segmented Hough histogram will
be computationally expensive.
In addition, the number of the contents of each bin will decrease,
which will reduce the contrast of the
histogram and make it difficult to determine the peak in the 
Hough histogram.

Instead of such an expensive analysis,
the discretized $\theta_{j}$ and $r_{j}$ values were corrected by fitting
Eq. (\ref{trackeq}) to the hit pixels in the track space.
For the correction,
only the middle points between the leading edges and the trailing edges
in the anode and cathode images were included in the fit.
The red hatched histogram in Fig.~\ref{rela_ang} shows the histogram of 
the opening angle between the two trajectories calculated from
the corrected $\theta$ values.
The standard deviation becomes 1.9$^{\circ}$,
which is about two times better than the one before the correction.
Assuming that the angular resolutions for two trajectories
are the same,
the angular resolution for one trajectory is calculated to be
$\sigma=\sigma(\theta_{3}+\theta_{4})/\sqrt{2}=1.3^{\circ}$.
This resolution agrees with the result from a Monte-Carlo simulation assuming the
in-plane angular resolution of 0.4$^{\circ}$ discussed in Section~\ref{secangreso}
and taking into account the error propagation in the three-dimensional
track reconstruction from the anode and cathode images.

Angular resolution achieved by the existing TPC based active targets is 
around 1$^{\circ}$ \cite{Roger2011,RANSAC}.
MAIKo demonstrates comparable performance to these active targets.

\section{Summary and outlook}
\label{secsummary}
An active target system MAIKo has been developed by Kyoto University and RCNP
for use in missing-mass spectroscopy
on unstable nuclei in inverse kinematics.
By detecting the recoil particles inside the detector volume,
measurements at forward angles are enabled.


The TPC was commissioned with the He+iso-C$_{4}$H$_{10}$ and
He+CO$_{2}$ gas mixtures at 430 hPa.
An electric field of about 200 V/cm was applied where the electron drift velocity was
about 1.5 cm/$\mu$s.
The maximum gas gain achieved by the $\mu$-PIC was about 3000.

The detector performances were examined by using a 56-MeV $^{4}$He beam at RCNP.
The angular resolution of the TPC for the beam particle was measured
under various beam intensities.
The TPC was successfully operated with the He+iso-C$_{4}$H$_{10}$ gas even when the
beam intensity was higher than 1000 kcps.
However, the operation was not stable at 100 kcps due to the discharge of the
$\mu$-PIC when the He+CO$_{2}$ gas was used.

Events in which the beam particles were scattered from the TPC gas were acquired.
A track finding algorithm based on the Hough transformation was developed.
The tracking algorithm was benchmarked by analyzing the $^{4}$He +$^{4}$He elastic
scattering events.
Multiple tracks were successfully separated into the individual tracks.
The angular resolution for the scattered particles was successfully
improved to 1.3$^{\circ}$ in sigma
by applying the fitting of the track after the Hough transformation;
the result is consistent with the in-plane angular resolution of $0.4^{\circ}$ determined
for beam particles.

Recently, the first measurement of the alpha inelastic scattering 
using an RI beam has been completed at RCNP.
A $^{10}$C secondary beam with an energy of 75 MeV/nucleon was injected into MAIKo.
The intensity of the beam was about 80 kcps.
According to the online analysis, the detection threshold for the recoil alpha particles
was less than 500 keV.
The detailed analysis is on going and the results will be reported elsewhere in the near future.

\section*{Acknowledgements}
The authors are grateful to the cyclotron crews at RCNP for providing the high-quality
and stable $^{4}$He beam.
The authors also thank Dr. H.~Baba from RIKEN for his valuable
advice on the data acquisition system.
T.~F. appreciates the JSPS Research Fellowship for young scientists under program
No. JP14J00949.
H.~J.~O. and I.~T. acknowledge the support of A.~Tohsaki and his spouse.
The present  work was supported by JSPS KAKENHI Grant Nos. JP20244030, JP23340068, 
JP23224008 and 15H02091.

\section*{References}

\bibliography{mybibfile}

\end{document}